\documentclass[trackchanges,twocolumn]{aastex701}

\usepackage{longtable}

\linenumbers

\makeatletter
\def\frontmatter@title@above{}
\renewcommand{\ps@apjheads}{%
  \let\@mkboth\markboth
  \def\@evenfoot{}\def\@evenhead{}
  \def\@oddfoot{} \def\@oddhead{}
}
\makeatother

\usepackage{amsmath} 

\definecolor{DanceGreen}{HTML}{054D33}
\definecolor{DanceLight}{HTML}{F2F6F4}
\definecolor{HeaderBlue}{HTML}{004488}
\definecolor{RowLightBlue}{HTML}{F0F6FA}

\usepackage[table]{xcolor} 
\usepackage{booktabs}      
\usepackage{multirow}      
\usepackage{graphicx}      

\renewcommand{\k}{\vec{k}}

\def\F{\mathcal{F}}

\def\S{\mathcal{S}}
\def\A{\mathcal{A}}
\def\B{\mathcal{B}}
\def\C{\mathcal{C}}
\def\N{\mathcal{N}}

\def\doi{http://doi.org}

\def\t{\tilde}
\def\e{\mathrm{e}}
\def\r{\mathrm{r}}
\def\g{\mathrm{g}}
\def\m{\mathrm{m}}

\def\s{\mathrm{s}}
\def\d{\mathrm{d}}

\renewcommand{\d}{\delta}

\newcommand{\HCd}{\mathcal{H}}

\def\HCdt0{\tilde{\HCd}_{0}}

\newcommand{\TNG}{IllustrisTNG}

\newcommand{\affbar}{Department of Physics, Faculty of Exact Sciences, Bar-Ilan University Ramat-Gan, 5290002 Israel}

\newcommand{\affDuke}{Department of Physics, Duke University, Durham, NC 27708, USA}

\newcommand{\affHel}{Helsinki Institute of Physics, P.O. Box 64, FI-00014 University of Helsinki, Finland}

\newcommand{\affSpace}{Space Telescope Science Institute, 3700 San Martin Drive, Baltimore, MD 21218, USA}
\newcommand{\affJhon}{Department of Physics and Astronomy, Johns Hopkins University, Baltimore, MD 21218, USA}

\begin{document}

\title{\large Debiasing Local Cluster Infall Reveals $\boldsymbol{H_0}$: Application to the Virgo Cluster}

\author[0000-0002-7234-844X]{Bastien Carreres}
\email[show]{bastien.carreres@duke.edu}
\affiliation{\affDuke}

\author[0000-0002-9578-3081]{David Benisty}
\email[show]{david.benisty@biu.ac.il}
\affiliation{\affbar}

\author[0000-0002-4999-3838]{Jenny Wagner}
\email[show]{\\ jenny.j.wagner@helsinki.fi}
\affiliation{\affHel}

\author{Avery Erlenbach}
\email[]{avery.erlenbach@duke.edu}
\affiliation{\affDuke}

\author[0000-0002-4934-5849]{Dan Scolnic}
\email[]{\\ daniel.scolnic@duke.edu}
\affiliation{\affDuke}

\author{Adam Riess}
\email[]{ariess@stsci.edu}
\affiliation{\affSpace}
\affiliation{\affJhon}

\begin{abstract}
Galaxy recession velocities are commonly used to measure the Hubble constant, $H_0$, with peculiar velocities treated as random noise. Around massive clusters, however, these motions can be predicted by the gravitational potential of the central halo, allowing them to be modeled rather than averaged over. Past analyses relied on analytic approximations for the distance-infall-velocity relation,  whose accuracy has not been tested using realistic cosmological simulations. We use halos from the IllustrisTNG simulation to quantify the bias and precision of several commonly used Hubble-flow models. We find that widely used fitting functions systematically underestimate $H_0$, whereas the full analytical solution reduces but does not eliminate these biases. We develop a simulation-based method to infer $H_0$ and apply it to TRGB distance samples around the Virgo Cluster. We find $H_0=74.4\pm 2.9$ and $73.4\pm 3.9$\, km\,s$^{-1}$\,Mpc$^{-1}$ ({\rm stat})  for the minor and the major infall models, respectively.
Correcting for the bias calibrated on TNG halos, these measurements raise the value of $H_0$ to $81.6 \pm 7.8$ and  $87.3 \pm 8.8$ km\,s$^{-1}$\,Mpc$^{-1}$ (stat+sys), where the minor infall model has the smallest bias from the Virgo mocks. We show that statistical errors are dominated by the residual velocity scatter around the infall model, while systematic errors are dominated by the transformation to the structure's comoving frame and halo population variance. These results establish a framework for interpreting cluster infall measurements and explain trends in previous $H_0$ inferences.
\end{abstract}


\section{Introduction}
\label{sec:introduction}
The present-day expansion rate of the Universe, quantified by the Hubble constant, $H_0$, is a central quantity in modern cosmology. Over the last decade, the tension between direct $H_0$ measurements in the local universe and measurements inferred from the early universe has reached the 5$\sigma $ level:
The local distance ladder anchored to Cepheid-calibrated Type~Ia supernovae yields $H_0 = (73.0 \pm 1.0) \, \mathrm{km\,s^{-1}\,Mpc^{-1}}$ \citep{bib:Riess2022}.
Likewise, a local distance network using a broad array of redundant distance indicators yields $H_0 = (73.5 \pm 0.8) \,\mathrm{km\,s^{-1}\,Mpc^{-1}}$ \citep{bib:Casertano2026}, while the cosmic microwave background (CMB) observations fitted to the cosmological concordance model, $\Lambda$CDM, yield
$H_0 = (67.19 \pm 0.38) \,\mathrm{km\,s^{-1}\,Mpc^{-1}}$ \citep{bib:Camphuis2026}.

This ``Hubble tension'' has persisted across a decade of improved data and has resisted simple explanations in terms of observational systematics, motivating the development of independent, complementary methods to constrain $H_0$, see, for instance \cite{bib:Aluri2023} or \cite{bib:Casertano2026} for recent summaries of the status-quo. Most of the complementary methods to measure $H_0$ locally consider different types of astronomical sources to independently test their uncertainties.

In contrast, the approach considered here infers $H_0$ from the interplay of the gravitational collapse of a galaxy cluster and the universe's expansion.
The application dates back to \cite{bib:Sandage1986}, who probed the Hubble flow around a spherical cosmic structure by observing that galaxies that fall into the structure are decelerated with respect to those that are only subject to cosmic expansion. 
At distances from the structure that are much larger than its extent, the structure's impact on galaxies is negligible.
These field galaxies only experience the background Hubble expansion, so their recession velocities increase linearly with their distance from the center of the structure.

At distances comparable to the structure's size, the gravitational attraction of the structure decelerates the flow of galaxies.
The zero-velocity radius, $r_0$, marks the distance from the structure's barycenter at which a galaxy's infall velocity toward the barycenter exactly cancels the Hubble recession. 
It separates galaxies that have already turned around and are falling in from those still moving outward.
Using $r_0$ as the extent of a cosmic structure today at redshift zero, probing the kinematic status of galaxies from the infall region to the free Hubble expansion allows one to jointly constrain the mass within $r_0$, $M(r_0)$, as well as $H_0$. 

While \cite{bib:Sandage1986} used a Newtonian solution for the recession velocity and distances that only effectively included $H_0$ as a linearly expanding background, \cite{bib:Peebles1984} had already included the Cosmological Constant, $\Lambda$, into the equations of motion and solved them numerically. 
The latter study also investigated different embedding cosmologies. 

For a fixed cosmology, the velocity-distance relation, or Hubble diagram, can be fitted to different distance indicators with a variety of calibrations. 
These include Cepheid-calibrated supernova distances, megamaser distances, and tip-of-the-red-giant-branch (TRGB) distances, such as those compiled in Cosmicflows-4~\citep{bib:Tully2023}. 
Within $\Lambda$CDM, the Hubble diagram has been used to infer $H_0$ or the mass for a growing number of systems: from galaxy groups~\citep{bib:Karachentsev2002,bib:Karachentsev2007,bib:Karachentsev2009,bib:Penarrubia2014,bib:Makarov2025,bib:Faucher2026,bib:Wagner2026} to galaxy clusters~\citep{bib:Karachentsev2010,bib:Nasonova2011,bib:Sorce2016,bib:Kim2020,bib:Benisty2026}. 
The functional form of the infall model employed has varied substantially across previous studies. The results for individual galaxy groups and galaxy clusters predominantly yield $H_0 \in [60, 80]~\mbox{km}\, \mbox{s}^{-1} \, \mbox{Mpc}^{-1}$ with uncertainties of the order of 10\%. 
Despite the comparably large uncertainties, the approach offers an alternative way to determine $H_0$ and to further investigate the source of the Hubble tension. 

Before applying any infall model to infer $H_0$, we need to test whether such a model accurately recovers the correct value of $H_0$ under realistic conditions. 
Previous work generally assumed that the models were accurate, without access to current, state-of-the-art simulations to test this assumption. 
The theoretical models generally describe an isolated, spherically collapsing structure embedded in a Friedmann background cosmology. 
We test the range of validity of these assumptions using the IllustrisTNG -- TNG300 \citep{bib:Nelson2019, Pillepich2018, Springel2018, Nelson2018, Naiman_2018, Marinacci2018} simulation, a large cosmological magneto-hydrodynamical simulation containing a large sample of galaxy-group- and -cluster-sized halos in a pre-defined $\Lambda$CDM universe.   
Once we determine the accuracy or bias in the method's application, we can apply it to the largest local cluster, Virgo.

The paper is organized as follows: in Section~\ref{sec:models}, we introduce the theoretical foundations and variants to model the Hubble flow. 
Section~\ref{sec:simulation} describes how we employ the IllustrisTNG halos to constrain $H_0$ via the theoretical models outlined in Section~\ref{sec:models} and to determine the bias and precision in $H_0$. A proof-of-concept for an observed structure is shown in Section~\ref{sec:virgo}, where we apply the Hubble flow fit to the Virgo galaxy cluster. Subsequently, Section~\ref{sec:discussion} summarizes the results. 


\section{Hubble Flow Models}
\label{sec:models}

A self-gravitating, bound system of objects embedded in an expanding cosmological background is often modeled as a pressure-less collection of ``particles" arranged in shells. 
These shells should not cross each other when they collapse under their own gravity while also being affected by the expansion of the background.
Under these conditions, the general relativistic equations yield the same description as their Newtonian analogue.
As detailed in Appendix~\ref{app:derivation_HFF}, there are three widely used approaches to model the Hubble diagram around a spherically collapsing structure like a galaxy group or a galaxy cluster: \citet{bib:Baushev2020}, \citet{bib:Peirani2008}, and \citet{bib:Penarrubia2014}. 

All three approaches share the assumptions stated above and model the radial infall velocity toward the center of the structure, $v(r)$, as a function of the physical distance from the center, $r$. 
For the zero-velocity radius, $r = r_0$, by definition, $v(r_0) = 0 \, \mbox{km} \, \mbox{s}^{-1}$, which implies that $v(r<r_0) < 0$ and $v(r>r_0) > 0$. 
The approaches also adopt similar initial conditions. 
The shell-forming particles initially co-move with the Hubble flow, reach a maximum radius, and then collapse under their own gravity. 
The volume of the structure within $r_\mathrm{vir}$ is assumed to be collapsed and virialized.
$r_\mathrm{vir}$ sets the smallest radius at which the collapse model without shell-crossing is valid. 
Therefore, the total enclosed mass $M(r_\mathrm{vir})$ within $r_\mathrm{vir}$ provides a lower bound on $M(r_0)$. 

\cite{bib:Baushev2020} treats each shell with $r > r_\mathrm{vir}$ as a collapsing Friedmann universe with enclosed mass density parameter $\Sigma_m$. $\Sigma_m$ is obtained by scaling the enclosed mass density in the shell with respect to the critical density of a flat $\Lambda$CDM, $\rho_\mathrm{crit}$, analogously to the density parameter of the background matter today, $\Omega_m$, which also equals $1 - \Omega_\Lambda$ in a flat $\Lambda$CDM. Even though the formalism allows for various $\Sigma_m$, we choose the ``point mass model" for computational efficiency. In this model, $\Sigma_m$ is the total mass density enclosed in the volume that all satellite galaxies jointly enclose, such that the mass resembles a point mass from the outside.

For a galaxy $g$ with $r_g > r_0$ and radial infall velocity $v_g$ today, the Hubble flow relation can be derived by equating the age of the background universe to the age of the overdensity today, at $t_0$.
The former is obtained from $\Lambda$CDM via $t_0 H_0 \sqrt{1-\Omega_m} = 2/3 \, \text{arcosh}\left(1/\sqrt{\Omega_m} \right)$.
We calculate the age of the collapsing structure from \cite{bib:Baushev2020}'s model, abbreviating $x\equiv r/r_g$: 
\begin{equation}
t_0 H_0 \sqrt{\Omega_\Lambda} = \int_0^1 \frac{\mathrm{d}x \sqrt{x}}{\sqrt{x^3+(U_g-\alpha-1)x+\alpha}} \;,
\label{eq:Baushev}
\end{equation}
with $\alpha \equiv \Sigma_m/\Omega_\Lambda$ and $U_g \equiv v_{g}^2/\left(r_{g}^2 H_0^2 \Omega_\Lambda\right)$.
Eq.~\eqref{eq:Baushev} implicitly defines a Hubble flow fit by relating the radial infall velocity $v_{g}$ and the observed distance $r_{g}$ of a galaxy to $H_0$ and the enclosed mass. 
The latter is calculated as $M = V_\mathrm{inf} \alpha (1-\Omega_m) \rho_\mathrm{crit}$ from the halo's mass density within the volume $V_\mathrm{inf}$. 
Galaxies with $r_\mathrm{vir} < r_{g} < r_0$ and $v_{g} < 0$ obey a similar implicit relation \citep{bib:Baushev2020}. 

\cite{bib:Peirani2006} and \cite{bib:Peirani2008} start from the same prerequisites. 
Instead of solving computationally intensive integrals for all galaxies, they set up an analytic fitting function as an approximation to the exact result that was only formulated in closed form years later in \cite{bib:Baushev2020}. The approximation reads: 
\begin{equation}
v(r) = - A \frac{H_0}{r^n}\left(\frac{G M}{H_0^2} \right)^{(n+1)/3} + B H_0 r \;,
\label{eq:Peirani}
\end{equation}
with $A=0.875$, $n=0.865$, and $B=1.274$ in \cite{bib:Peirani2006} and $A=0.976$, $n=0.627$, and $B=1.377$ in \cite{bib:Peirani2008}.
More recently, \cite{bib:Benisty2026a} fitted for these parameters, assuming $H_0$ and $M$ known in simulated data.  
They found that $A$ and $B$ were close to unity, while $n$ could not be constrained by these simulated halos.
Only at $r=r_0$ does the fitting function correspond to the numerical, exact solution of the collapse equations and $M$ is $M(r_0)$. 

\cite{bib:Penarrubia2014} assumes a point mass $M$ embedded in an expanding universe and Keplerian orbits around the point mass.
Despite the different ansatz, they set up their approximate fitting function to the Hubble flow also as Eq.~\eqref{eq:Peirani}.
Yet, they use $A=1.1$, $n=0.5$, $B=1.494$ as coefficients.
Modeling the enclosed mass as a point mass requires the orbiting galaxies to be distant enough, such that the point mass is a good approximation to the enclosed mass, i.e.~the enclosed mass is relaxed or an inner merger does not alter the satellite orbits in the structure's outskirts. 
An example of such a merger is the one in the Local Group, see \cite{bib:Penarrubia2014} for limits on the validity of the approximation.

\section{Application to simulated halos}
\label{sec:simulation}

In our analysis, we first compare the three approaches by \cite{bib:Baushev2020}, \cite{bib:Peirani2006}, and \cite{bib:Penarrubia2014} to simulated data in which the infall velocities are known and directly available. 
The evaluation shows how well realistically simulated structures can be approximated via the assumptions of the spherical collapse models of Section~\ref{sec:models}.

We use the $z=0$~snapshot of the IllustrisTNG300 simulation \citep{Pillepich:2019bmb}, part of the IllustrisTNG project \citep{bib:Vogelsberger2014,bib:Nelson2015,Nelson:2019jkf}.
IllustrisTNG300 is the largest IllustrisTNG volume with a periodic box of side length $205\, h^{-1} \, \mbox{Mpc}$, $h=0.6774$. 
Halo masses span $M_{200} \sim 10^{11}$-$10^{15}\,M_\odot$.

\subsection{Halo Selection}

We identify a halo as isolated if there is no neighboring halo with mass greater than or equal to $0.1 \, M_{200}$ within $r_0$, as given by \cite{bib:Sandage1986}: $r_0 = \left(8 t_0^2 G M/\pi^2 \right)^{1/3}$. As stated in \cite{bib:Lovell2018}, $M_{200}$ in the simulation is the total enclosed mass in a volume that has a mass density corresponding to 200 times the cosmic critical density. It thus corresponds to an enclosed mass $M(r_{200}) \approx M(r_\mathrm{vir})$ in the models of Section~\ref{sec:models}.

For each sub-halo~$g$ flagged as ``galaxy" with a mass larger than $10^9 \, M_\odot$, we obtain its
distance from the main halo center as
$r_g = |\mathbf{r}_g| = |\mathbf{r}_\mathrm{sub} - \mathbf{r}_{\rm c}|$, with $\mathbf{r}_\mathrm{sub}$ and $\mathbf{r}_{\rm c}$ being the three-dimensional positions of the sub-halo and the center of the main halo, respectively, and $|\cdot|$ the Euclidean norm of a vector.
The sub-halo's radial infall velocity towards the main halo center is obtained by 
$v_g = (\mathbf{v}_\mathrm{g}-\mathbf{v}_\mathrm{c}) \cdot \mathbf{r}_g/r_g + H_0^{\rm true} r_g$, in which $\mathbf{v}_\mathrm{g}$ and $\mathbf{v}_\mathrm{c}$ are the three-dimensional velocities of the sub-halo and the main halo as given by the simulation. 

\begin{figure}[htbp]
    \centering
\includegraphics[width=0.95\linewidth]{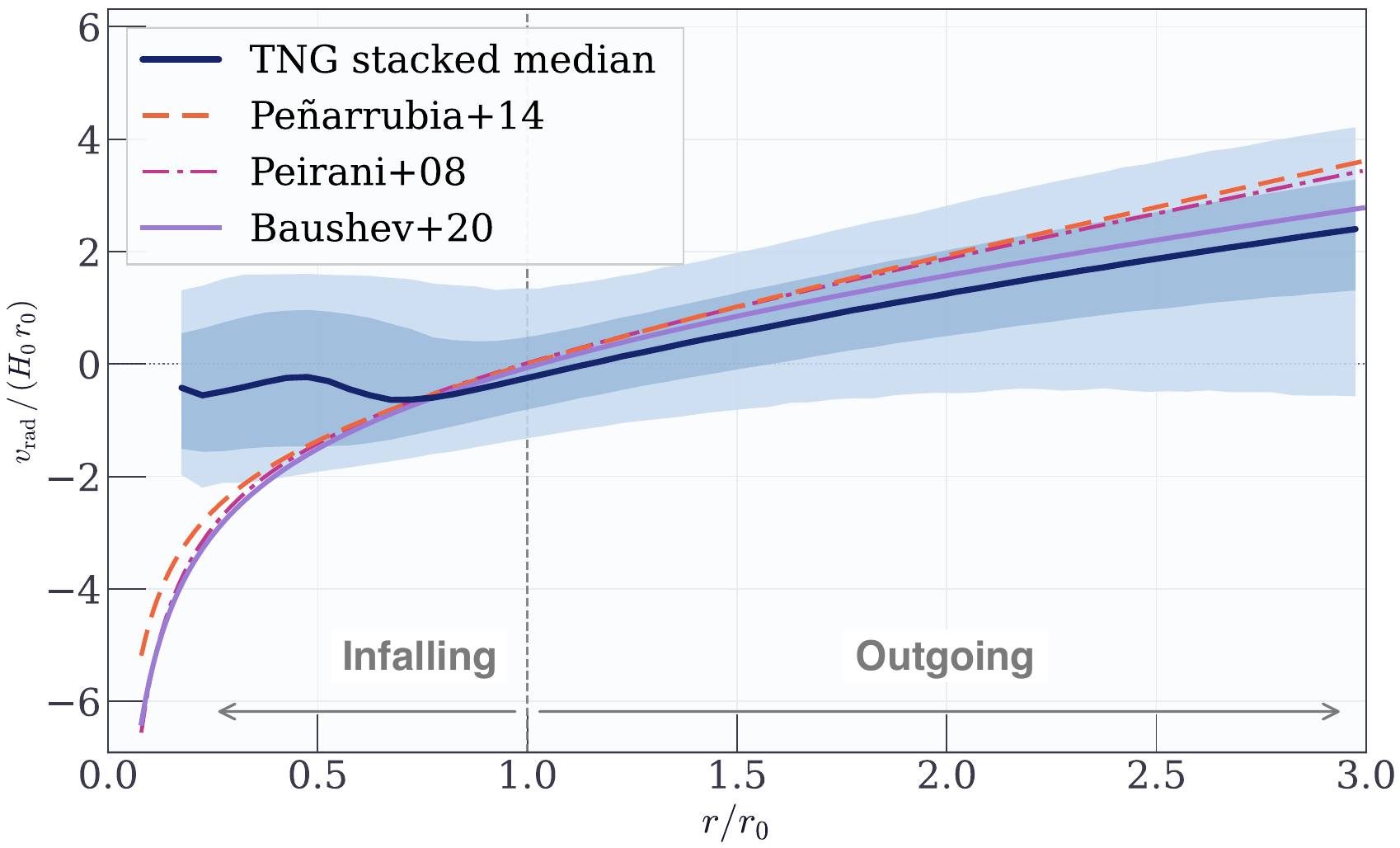}
\caption{Stacked median radial velocity profile of isolated \TNG300\ halos ($M_{200}\geq10^{12}\,M_\odot$), normalized by $H_0 r_0$ and $r_0$ (dark blue line) with 1- and 2-$\sigma$ bands of standard deviation of the sample (blue shaded regions). Hubble flow fits by \citet{bib:Peirani2008} (dash-dotted magenta line, $v\to1.377\,H_0 r$), \citet{bib:Penarrubia2014} (dashed yellow line, $v\to1.494\,H_0 r$), \citet{bib:Baushev2020} exact solution (solid purple line, $v\to H_0 r$).}
\label{fig:hubble_flow_stacked}
\end{figure}

Fig.~\ref{fig:hubble_flow_stacked} shows the median Hubble diagram stacking all TNG300 halos with $M_{200}\geq10^{12} \, M_\odot$.
We compare the median Hubble diagram with the three fitting functions introduced in Section~\ref{sec:models}, setting the mass parameter, $M$, in each model to $M(r_\mathrm{min})$ with $r_\mathrm{min}$ being the radius of the first sub-halo used in the Hubble flow fit. 
In addition, we show the 1- and 2-$\sigma$ bands of standard deviation of the halo population around the median.
After normalization, all curves agree at $r=r_\mathrm{min}$. 
Eq.~\eqref{eq:Baushev} follows the median Hubble flow for $r \, \ge r_\mathrm{min} \approx \, 0.7 \, r_0$, suggesting that the simulated structures can be approximated by the spherical collapse assumptions made in \cite{bib:Baushev2020} within the band of 1-$\sigma$ standard deviation.
As Eq.~\eqref{eq:Peirani} is an approximation, its deviation from the median curve grows at smaller and larger $r/r_0$, see also Section~\ref{sec:discussion}. 
Fig.~\ref{fig:hubble_flow_stacked} shows similar curves for Eq.~\eqref{eq:Peirani} with the coefficients of \cite{bib:Peirani2008} and \cite{bib:Penarrubia2014}.
Due to these similarities, in the rest of the analysis, we decide to only investigate the two fits of \cite{bib:Penarrubia2014} and \cite{bib:Baushev2020}.  

\subsection{Hubble Flow Fitting}

With these prerequisites, for each sub-halo $g$, 
\begin{equation}
    \chi_g^2 = \left(\frac{v_g - v_{\rm fit}(r_g; H_0, \log M)}{\sigma_g(H_0, \log M)}\right)^2 \;,
\end{equation}
in which $v_\mathrm{fit}$ is a Hubble flow fitting model depending on $H_0$ and $M$ from Section~\ref{sec:models} and $\sigma_g$ is defined by
\begin{equation}
    \sigma_g^2(H_0, \log M) = \sigma_{v, \text{meas}}^2 + \left(\left.\frac{\partial v_{\rm fit}}{\partial r}\right|_{r=r_g}\sigma_{r_g}\right)^2 + \sigma_v^2 \;.
\end{equation}
We assume a velocity measurement uncertainty of $\sigma_{v, \text{meas}} = 3$~km~s$^{-1}$%
for all sub-halos corresponding to the typical spectroscopic measurement uncertainty and a distance uncertainty of $\sigma_{r_g} = 0.05 \, r_g$, representative of the currently most precise distance indicators. 
We include a free constant parameter, $\sigma_v$, as done in \cite{bib:Penarrubia2014} and  \cite{bib:Kim2020}, to allow for an overall velocity scatter.
A systematic investigation of its impact and alternatives to model these effects for the Local Group as a cosmic structure can be found in Wagner et al. (in prep.).

We then define the likelihood of each galaxy as
\begin{equation}
   \mathcal{L}_g =\frac{1}{\sqrt{2\pi \sigma_g^2}}\exp\left(-\frac{1}{2}\chi_g^2\right)
\end{equation}
and maximize the total likelihood, $\mathcal{L} = \prod_{g=1}^{n_\mathrm{sub}} \mathcal{L}_g$, for all $n_\mathrm{sub}$ sub-halos of a main halo in the fit to obtain $H_0$ and $M$. 
Introducing $\sigma_v$, the fit reduced $\chi^2$ will tend to 1 by construction of the optimization, such that the amplitude of $\sigma_v$, that is a proxy for the scatter of velocities around the fitted infall model, can be interpreted as an indicator for the quality of the fit.

We only use halos with at least $n_\mathrm{sub}=10$ sub-halos. 
The latter lie at distances $0.8 \, r_0 < r_g < 2.5 \, r_0$ from the halo center. Boundaries on $H_0$ and $\log M$ are chosen as $(10<H_0<150) \, \mathrm{km\,s^{-1}\,Mpc^{-1}}$ and $11<\log_{10} (M/M_\odot)<16$. 
Further implementation details and boundary tests are given in Appendix~\ref{app:implementation}. 

\begin{figure*}
\centering
\includegraphics[width=\textwidth]{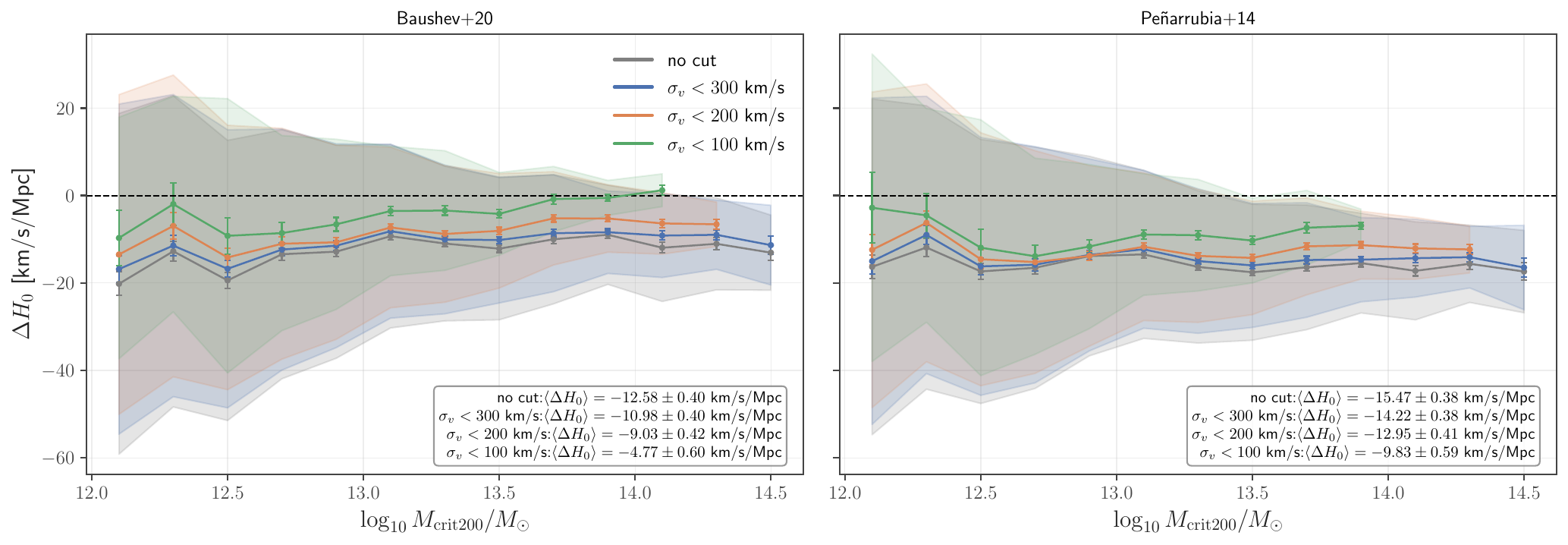}
\caption{$H_0$-bias $\Delta H_0\equiv H_0-H_0^{\rm true}$ versus true halo mass for the \citet{bib:Baushev2020} Hubble flow fit (left) and the \citet{bib:Penarrubia2014} fit (right) in $0.8 \, r_0<r<2.5 \, r_0$. Points show the per-bin mean, error bars the 1-$\sigma$ standard deviation of the mean, and shaded areas indicate the 1-$\sigma$ standard deviation of the sample. $\sigma_v$ is the free hyperparameter in the uncertainty of the Hubble flow fit to absorb an overall scatter around the Hubble flow serving as quality-of-fit indicator (see text).} 
\label{fig:H0_bias}
\vspace{2ex}
\includegraphics[width=\textwidth]{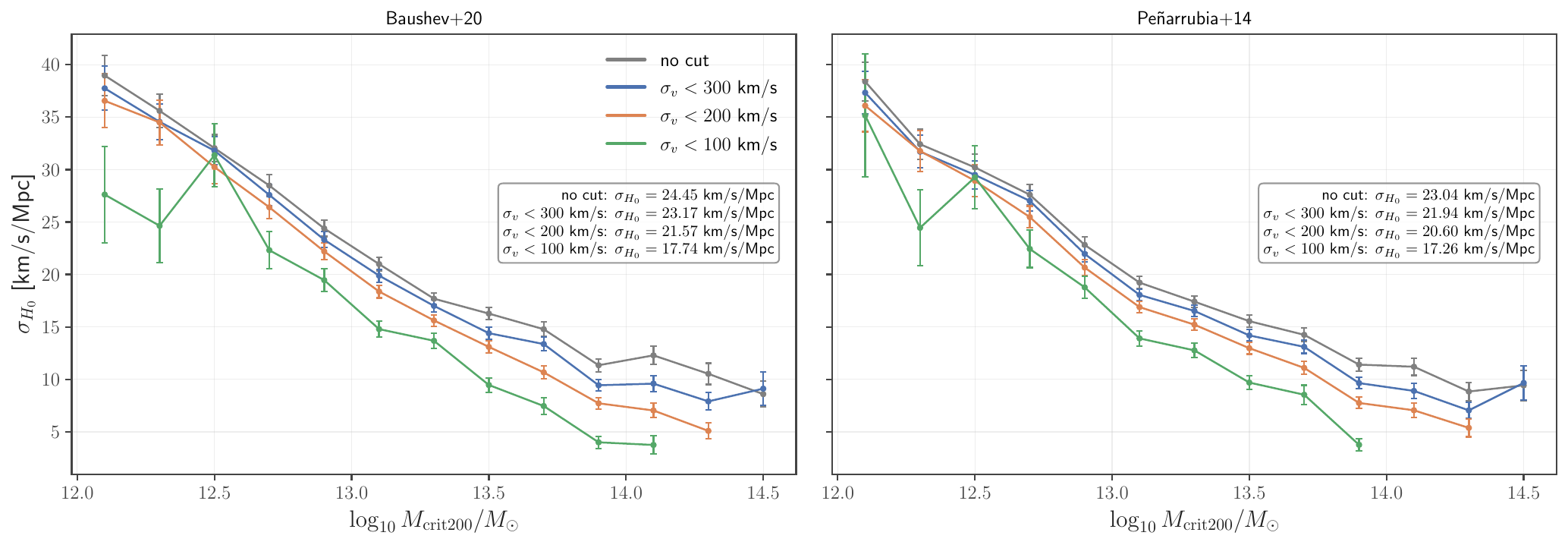}
\caption{Standard deviation of the halo population in the inferred $H_0$, $\sigma_{H_0}$, versus true halo mass for the \citet{bib:Baushev2020} Hubble flow fit (left) and the \citet{bib:Penarrubia2014} fit (right) in $0.8 \, r_0<r<2.5 \, r_0$. Binning, colors and error bars are the same as in Fig.~\ref{fig:H0_bias}.}
\label{fig:H0_scatter}
\end{figure*}

\subsection{Results}

Our results only contain halos with converged fits and a valid error estimation according to the \texttt{iminuit} fitting algorithm that we used. 
In addition, we select fits with a $p$-value larger than $0.01$ to remove outliers which cannot be approximated by the spherical collapse models, see also Fig.~\ref{fig:qof} in Appendix~\ref{app:implementation} for some examples.

Fig.~\ref{fig:H0_bias} shows the bias in the $H_0$-value as a function of the true halo mass, $M_{200}$, for the \cite{bib:Baushev2020} model (left) and the \cite{bib:Penarrubia2014} model (right), depending on the $\sigma_v$ parameter. 
We bin the results by true halo mass using bins of width $\Delta \log_{10} (M / M_\odot) = 0.2$.
The shaded regions are the standard deviation of the sample (population of halos), $\sigma_\mathrm{halo}$, when determining $H_0$.
The error bars in each bin mark the uncertainty of the bias, i.e.~the standard deviation of the mean, $\sigma_\mathrm{halo}/\sqrt{n_\mathrm{halo}}$.

For halos with $M_{200}>10^{13}\, M_\odot$, the non-zero bias in $H_0$ becomes approximately independent of the binned halo mass, clearly showing the limits of the oversimplified spherical collapse models when applied to realistic structures.
Cluster-scale $H_0$-inferences have a greater statistical precision than group-scale inferences.
The reason for this lies in their deeper gravitational potential that reduces the perturbing impact of smaller-scale halos in the environment and the increased number of sub-halos improving the precision of the fit.
Inferences of $H_0$ from galaxy groups with $M_{200} \lesssim 10^{13}\,M_\odot$ should therefore be treated with caution.

\begin{figure}[t!]
    \centering
\includegraphics[width=0.85\linewidth]{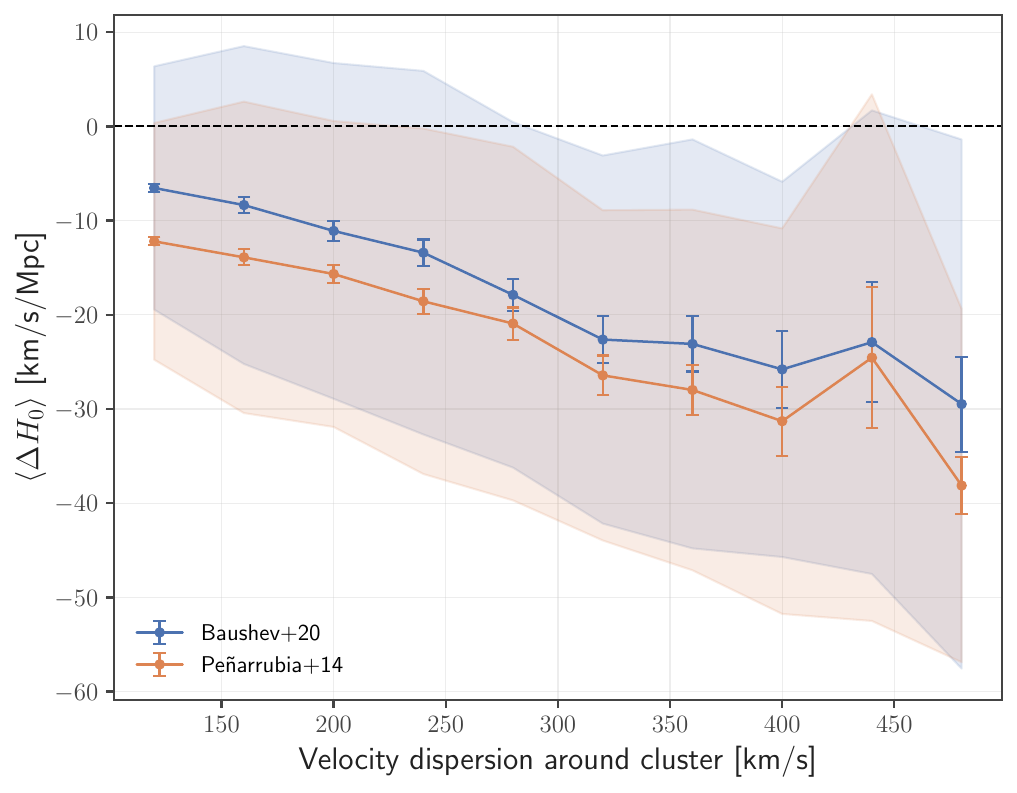}
\caption{Average $H_0$-bias, $\langle \Delta H_0 \rangle$, as a function of the velocity dispersion around clusters of mass $M>10^{13}M_\odot$ for the fitted value from the \cite{bib:Baushev2020} model (blue) and the \cite{bib:Penarrubia2014} model (orange).} 
\label{fig:std_vel}
\end{figure}

The $H_0$-bias could be linked to aspherical shapes and anisotropic galaxy accretion by non-relaxed, late-formed clusters, see \cite{bib:Gouin2021} and the impact of additional interlopers and substructures distorting the Hubble flow of a halo. 
Halos with low $\sigma_v$ and a correspondingly low $\sigma_{H_0}$ might be early-formed, relaxed ones that are rather isolated and therefore match the criteria set up in the Hubble flow models \citep{bib:Gouin2021}. 
Yet, only for the \cite{bib:Baushev2020} model, the bias approaches zero for high mass halos and $\sigma_v < 100$~km~s$^{-1}$.
So, for these IllustrisTNG300 halos, \cite{bib:Baushev2020} provides a more accurate approximation than \cite{bib:Penarrubia2014}.
The result is also supported by Fig.~\ref{fig:std_vel}, which shows the averaged bias versus the velocity dispersion of the Hubble flow diagram.
Yet, the plots also show that $\sigma_v$ is a necessary parameter and the spherical collapse model alone does not capture all features of a realistic halo.

\section{Application to the Virgo Cluster}
\label{sec:virgo}

Having established how well simulated galaxy clusters satisfy the requirements for a Hubble flow fit of a spherical collapse model, we demonstrate the application of the Hubble flow fit to observations.
As Section~\ref{sec:simulation} revealed, on average in a $\Lambda$CDM cosmology, cluster-scale structures show a higher precision when inferring $H_0$ compared to group-scale structures.
Therefore, we analyze the nearest galaxy cluster, the Virgo galaxy cluster.

In contrast to the simulated halos that provide all information required by the collapse models, only line-of-sight velocities measured relative to the Solar System are known for most observed structures. 
These line-of-sight velocities contain the radial infall velocity of a galaxy towards the structure's barycenter, but also a velocity component due to the galaxy's motion in the potentially perturbed environment, and even a velocity component of the observer that need not be co-moving with the structure.
The radial infall velocities required by the models must be reconstructed from these measurements, as described, for example, by \cite{bib:Karachentsev2006} and \cite{bib:Wagner2025}. 
The reconstruction comes at the cost of additional uncertainties and further model assumptions whose impacts we show here.

\subsection{Data Selection}
\label{sec:virgo_data}
Virgo consists of at least three major parts, the largest of which is centered on the galaxy M~87 (NGC~4486) at SGL=102.88$^\circ$ and SGB=-2.35$^\circ$.
Along this direction, estimates of the distance to the cluster barycenter range from 15 to 17~Mpc. 
Estimates of its group velocity range from 900 to 1100~km/s \citep{bib:Kashibadze2018,bib:Kashibadze2020,bib:Anand2025,bib:Makarov2026,2026arXiv260718414F}.
Following \cite{bib:Kim2020}, we adopt $d_\mathrm{C}=(16.5 \pm 0.1) \, \mbox{Mpc}$ for our distance to the Virgo barycenter and $v_\mathrm{C}=(988 \pm 61) \, \mbox{km}\, \mbox{s}^{-1}$ for its group velocity in the Local-Group frame.

The most recent Hubble flow analysis by \cite{bib:Kim2020} fitted 33 galaxies with TRGB distances selected in a cone around our line of sight.
We instead collected two larger samples. 
The first contains 107 galaxies with TRGB-based distances compiled by \cite{bib:Anand2021} and \cite{bib:Anand2025} with velocities taken from \cite{bib:Tully2023}.
The second sample contains 71 galaxies with TRGB-based distances compiled from \cite{bib:Kashibadze2018} and \cite{bib:Kashibadze2020}, see Tables~\ref{tab:virgo_Anand} and \ref{tab:virgo_Kashibadze} in Appendix~\ref{app:data_virgo}.

Due to the differing calibrations of both datasets, it is not clear a priori whether they can be combined into a single dataset. 
Therefore, we first analyze them separately to compare their inferred Hubble flow characteristics to confirm that we can join them. 
The joint dataset consists of 121 galaxies that are taken from Table~\ref{tab:virgo_Anand} and filled up with the missing ones from Table~\ref{tab:virgo_Kashibadze}.

In an analogous comparison, \cite{bib:Kim2020} found that changes in the zero point and in the relation between TRGB colors and the F814W TRGB magnitudes did not produce a bias in $H_0$ relative to the uncertainties of their dataset. 
As shown by \cite{bib:Anand2025}, the TRGB- and SBF-based distance estimates agree within their uncertainties.
We therefore retain the SBF-distance-based fiducial value of $d_\mathrm{C}$ in the Hubble flow fit.

\subsection{Data Preprocessing via Infall Models}
\label{sec:virgo_preprocessing}

For each galaxy, we first convert its line-of-sight velocity into the Local-Group rest frame using the same values and definitions as in \cite{bib:Wagner2026}.  
Then, we use these Local-Group-frame line-of-sight velocities to approximate the radial infall velocities with respect to the barycenter of Virgo with the minor and major infall model velocities \citep{bib:Karachentsev2007}.
As detailed by \cite{bib:Wagner2025}, the two infall models approximate a galaxy's radial infall velocity relative to the barycenter of a cosmic structure when only its line-of-sight component of the velocity vector is known and proper motion information is unavailable. 

Both models make the best use of the line-of-sight velocity information when the angle $\lambda$ between the observer, the infalling galaxy, and the barycenter of the cosmic structure satisfies $|\cos \lambda | \approx 1$. 
In the minor infall model, galaxies with $\cos \lambda =0$ contribute no radial-velocity information, so that a bias is introduced if no angular cut for these galaxies is applied. 
By contrast, the major infall model diverges at $\cos \lambda =0$, making such cuts necessary. 
This is why cuts on $\lambda$ are often imposed, e.g.~in \cite{bib:Karachentsev2010} or \cite{bib:Kim2020}. 
Although skipping angular cuts can bias the infall-model velocities, applying cuts does not necessarily improve their accuracy because the models neglect non-zero perpendicular velocity components \citep{bib:Wagner2025}. 
This use of infall models is the first difference to the simulation analysis in Section~\ref{sec:simulation}, in which the full three-dimensional velocity information was available. 

The second difference is the completeness of the Hubble flow galaxies.
Our samples are necessarily incomplete and subject to selection effects because we preferentially observe galaxies on our near side of Virgo and we rely on a single distance indicator, see \cite{bib:Kashibadze2018} and \cite{bib:Kashibadze2020} for further details. 
For the expected range of $r_0 \in [5,7]\, \mbox{Mpc}$ \citep{bib:Kashibadze2018}, our datasets with distances less than 15 Mpc from the Virgo barycenter extend to more than $2 \, r_0$.
After excluding galaxies within 4~Mpc of the Virgo barycenter, the datasets begin approximately at $0.6\, r_0$ to $0.8 \, r_0$.

\subsection{Mock Observation Inferences}
\label{sec:virgo_mock}

To identify and quantify potential biases of the Virgo Hubble flow fit, we refine the simulation-based validation of Section~\ref{sec:simulation}: 
we select Virgo-like halos as those with $M_{200} >10^{14}~M_\odot$, which is a lower bound based on X-ray observations. 
Then, we place mock observers at positions and relative velocities, $\hat{d}_\mathrm{C}$, $\hat{v}_\mathrm{C}$, around the halos with $\hat{d}_\mathrm{C} \sim \mathcal{N}(16.5\, \mbox{Mpc},0.1\, \mbox{Mpc})$ sampled from a Gaussian distribution. 
For $\hat{v}_\mathrm{C}$, the mean value is calculated as:
\begin{equation}
\tilde{v}_\mathrm{C} = H_0^\mathrm{true} d_\mathrm{C} + \boldsymbol{v}_\mathrm{c} \cdot \frac{\boldsymbol{r}_\mathrm{C}-\boldsymbol{r}_\mathrm{obs}}{|\boldsymbol{r}_\mathrm{C}-\boldsymbol{r}_\mathrm{obs}|} \;,
\label{eq:vcdc}
\end{equation}
in which the last term represents the projection of the Virgo-like halo's velocity onto the unit vector from the observer towards the halo's barycenter.
Based on this, $\hat{v}_\mathrm{C}\sim \mathcal{N}(\tilde{v}_\mathrm{C}, 61\, \mbox{km} \, \mbox{s}^{-1})$.
The random positioning is supposed to mimic the effect of the uncertain $d_\mathrm{C}$ and $v_\mathrm{C}$ in the observables.
Then, the sub-halos are also transformed to the mock-observation frame and realistic 5\% uncertainties on their distances and $3\, \mbox{km} \, \mbox{s}^{-1}$ uncertainties on the line-of-sight velocities are added as in Section~\ref{sec:simulation}. 
We apply the infall models to these mock observations in the same way as for the observations.

The Hubble flow fit using \cite{bib:Baushev2020}'s model as detailed in Section~\ref{sec:simulation} assumed $d_\mathrm{C}$ and $v_\mathrm{C}$ to be known. The observables cannot provide this information, so that the fit needs to be amended: the independent barycenter estimates from the early-type galaxies in the cluster core are used as priors in an extended Hubble flow fit that infers $d_\mathrm{C}$- and $v_\mathrm{C}$-values as well. 
During that fit, we center Gaussian priors of $d_\mathrm{C}$ and $v_\mathrm{C}$ on $\hat{d}_\mathrm{C}$ and $\hat{v}_\mathrm{C}$ with the corresponding error scales. 
Due to the degeneracy between the barycenter parameters and $H_0$, see Eq.~\eqref{eq:vcdc}, we additionally perform a three-parameter Hubble flow fit, as done in Section~\ref{sec:simulation}, in which 
the uncertainties in $d_\mathrm{C}$ and $v_\mathrm{C}$ enter in the infall models, such that the distances and the approximated infall velocities incorporate the uncertainties.
The resulting Hubble flow parameters are then conditioned on a fiducial set of $d_\mathrm{C}$ and $v_\mathrm{C}$.
In all fits, we select a subsample of ``good-fits'' by removing pathological fits flagged by \texttt{iminuit}, cutting on the $p$-value $p > 0.01$ and requiring an inferred $\sigma_v < 200 \, \mbox{km} \, \mbox{s}^{-1}$.

\begin{figure}
    \centering
    \includegraphics[width=0.95\columnwidth]{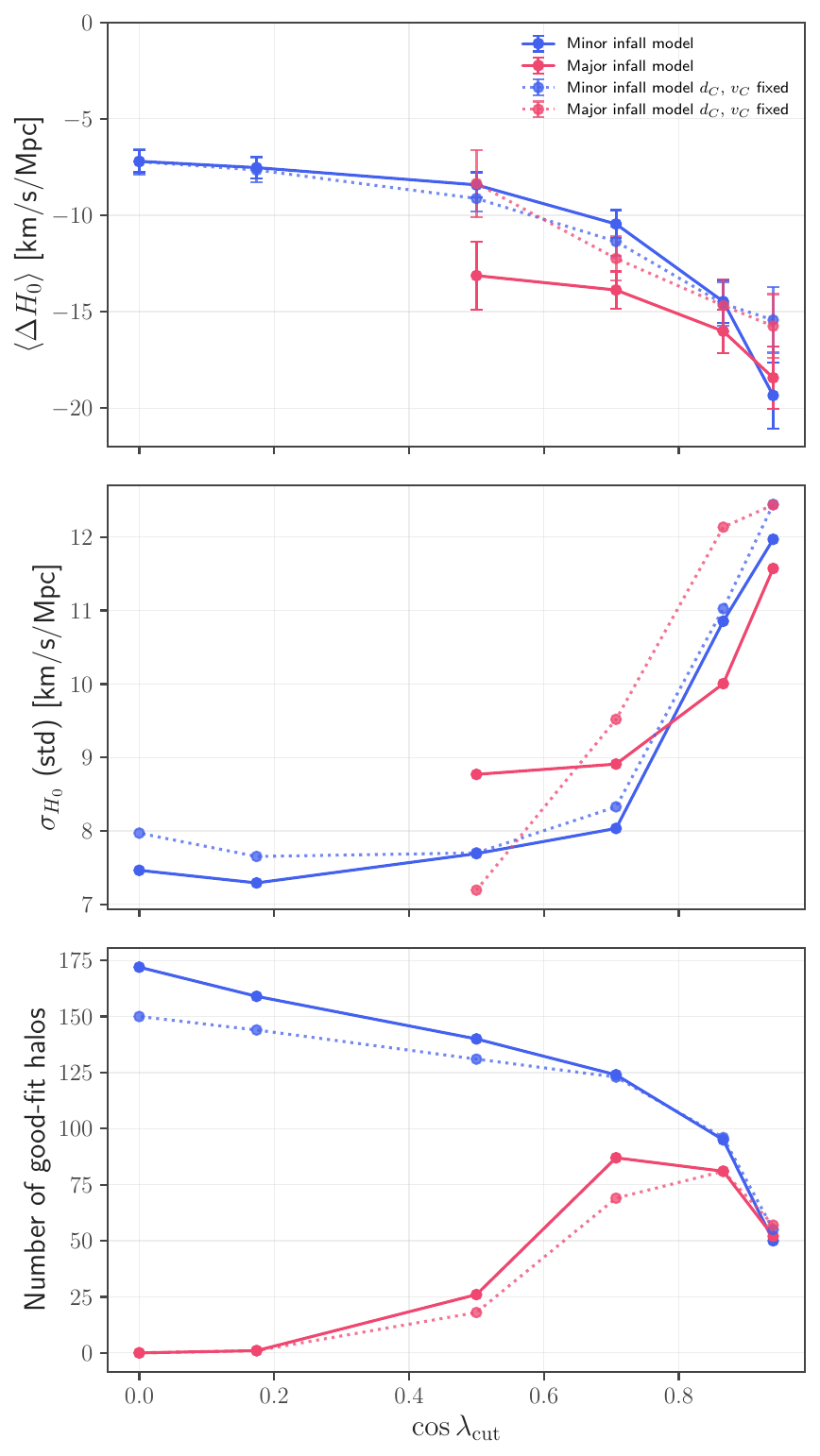}
    \caption{Impact of $\lambda$-cuts on the $H_0$-bias and quality-of-fit for Virgo-like mocks. Mean bias of $H_0$ as a function of the cut on the infall angle, using only $|\cos \lambda| > \cos \lambda_\text{cut}$ with $0^\circ \le \lambda_\mathrm{cut} \le 90^\circ$, (top plot), standard deviation of the sample in $H_0$ (center plot), and the number of halos flagged as ``good-fit'' (bottom plot) for the minor (blue line) and major infall model (red).  Solid lines show baseline result when $d_\mathrm{C}$ and $v_\mathrm{C}$ are fitted using a prior (5-parameter fit) and dotted lines show result when $d_\mathrm{C}$ and $v_\mathrm{C}$ are fixed (3-parameter fit).}
\label{fig:lambdacut}
\end{figure}

In Fig.~\ref{fig:lambdacut}, we show the impact of selecting sub-halos according to their infall angle $\lambda$ on the $H_0$-bias. The cut is defined such that $|\cos\lambda| > \cos\lambda_\text{cut}$.
The upper plot shows the bias in $H_0$ averaged over all halos.
It reveals that the bias decreases for both the minor and the major infall models the fewer sub-halos are cut out.
For the minor infall, the smallest bias is attained when including almost all sub-halos, i.e.~$\cos \lambda_\mathrm{cut} \simeq 0.17$, and remains relatively stable when all sub-halos are included. For the major infall, the $H_0$-bias is larger than for the minor infall model, as are the uncertainties (error bars) on the inferred value.

The standard deviation of the sample in $H_0$, $\sigma_{H_0}$, in the center plot, follows this trend.
For example, the precision in $H_0$ is increased by roughly 60\% comparing the tightest to no $\lambda$-cuts.
For the major infall model, the bottom plot showing the number of ``good-fit'' halos indicates that the major infall model should have a cut obeying $|\cos \lambda| \ge \cos \lambda_\mathrm{cut} \approx 0.707$ to maximize the number of halos in the statistics. 

We can see that, for the minor infall model, choosing the 5- or the 3-parameter fitting approach does not significantly impact those results, but leads to about 15\% fewer ``good-fits'' when $\cos\lambda_\mathrm{cut} \rightarrow 0$. 
For the major infall model, fixing $d_\mathrm{C}$ and $v_\mathrm{C}$ leads to a lower bias but mostly to a higher scatter. 
We notice that, due to its pathological behavior when $\cos\lambda \rightarrow 0$, the number of ``good-fit'' halos falls to zero after a cut of $\cos\lambda_\text{cut} \simeq 0.17$. 

As the minor infall model yields a smaller $H_0$-bias at a simultaneously lower $\sigma_{H_0}$ for a larger number of well-fitted halos, it is preferred over the major infall model based on the mocks.

Thus, extending the validation of Section~\ref{sec:simulation} to account for the missing velocity components in observations, we find that the often-used $\lambda$-cuts actually worsen the bias when inferring $H_0$.
So the argument that the cuts are implemented to optimally use the line-of-sight velocity information turns out to be invalid, at least for the simulation employed here.

\begin{figure}[t!]
\includegraphics[width=0.95\linewidth]{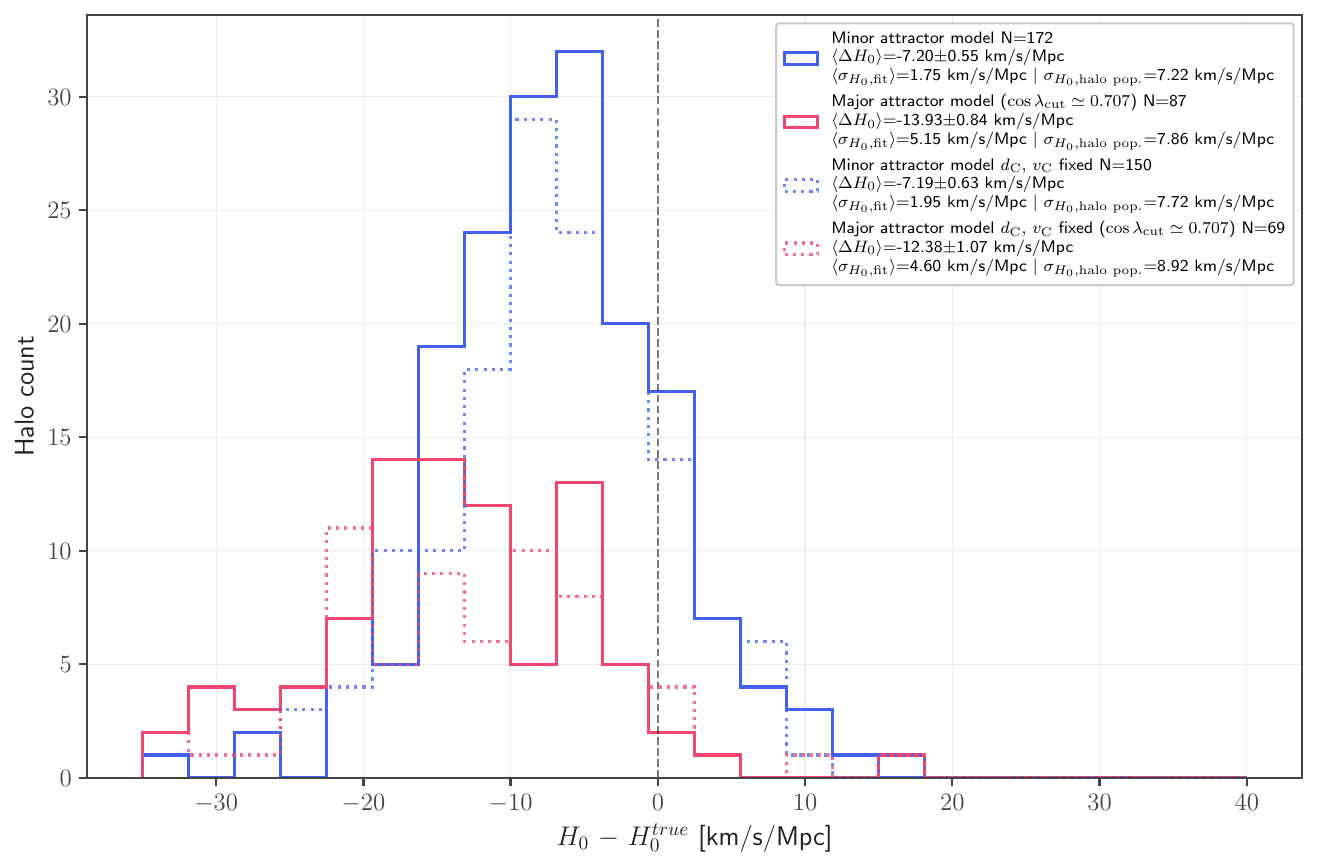}
\caption{Distribution of the $H_0$-bias from the Virgo-like subsample of the simulation, $M_{200}>10^{14}\, M_\odot$, using a major infall model with $| \cos \lambda | > 0.707$ (red) and a minor infall model without cuts (blue). Solid lines show result when $d_\mathrm{C}$ and $v_\mathrm{C}$ are fitted using a prior and dotted lines show result when $d_\mathrm{C}$ and $v_\mathrm{C}$ are fixed.}
\label{fig:mock_lambda_cuts}
\end{figure}

Fig.~\ref{fig:mock_lambda_cuts} summarizes the distribution of $H_0$-values inferred from all halos for the minor infall model without cuts and the major infall model with the optimum $\lambda$-cut for the 5- and 3-parameter Hubble flow fits.
Comparing the standard deviations of all individual Hubble flow fits with the standard deviation of the sample for both infall models, we note that the former is smaller than the latter. 
This implies that the population includes additional variances that are not taken into account in the modeling. 
As noted earlier, these variances mark the limits of the spherical collapse model and may be reduced either by a better characterization of the sub-population of halos for which the model's assumptions more closely apply, or by an improved collapse model.
Overall, in the 5-parameter fit, we obtain a bias $b_{H_0} = (-7.2 \pm 0.6) \, \mbox{km} \, \mbox{s}^{-1} \, \mbox{Mpc}^{-1}$ with $\sigma_{H_0} = 7.2\, \mbox{km} \, \mbox{s}^{-1} \, \mbox{Mpc}^{-1}$ for the minor infall model and $b_{H_0} = (-13.9 \pm 0.8) \, \mbox{km} \, \mbox{s}^{-1} \, \mbox{Mpc}^{-1}$ with $\sigma_{H_0} = 7.9\, \mbox{km} \, \mbox{s}^{-1} \, \mbox{Mpc}^{-1}$ for the major infall model. The bias for the 3-parameter model is within the error bounds of the 5-parameter fit, as can be read off from Fig.~\ref{fig:mock_lambda_cuts}.

It remains an open question whether these mock observations adequately represent an individual structure like the Virgo cluster.
The choices we made when selecting comparable halos may not cover the full range of possible variation or may be too broad. 
We also found that the bias is sensitive to the $\sigma_v$ quality cuts and may be further reduced by imposing stricter cuts.
Hence, further work is necessary to determine if the $H_0$-bias can be used to de-bias the observationally-inferred $H_0$ of the Virgo cluster or, more generally, of an individual cosmic structure.

\subsection{Results from observations}
\label{sec:virgo_results}
\begin{figure*}[t!]
    \centering
\includegraphics[width=0.85\linewidth]{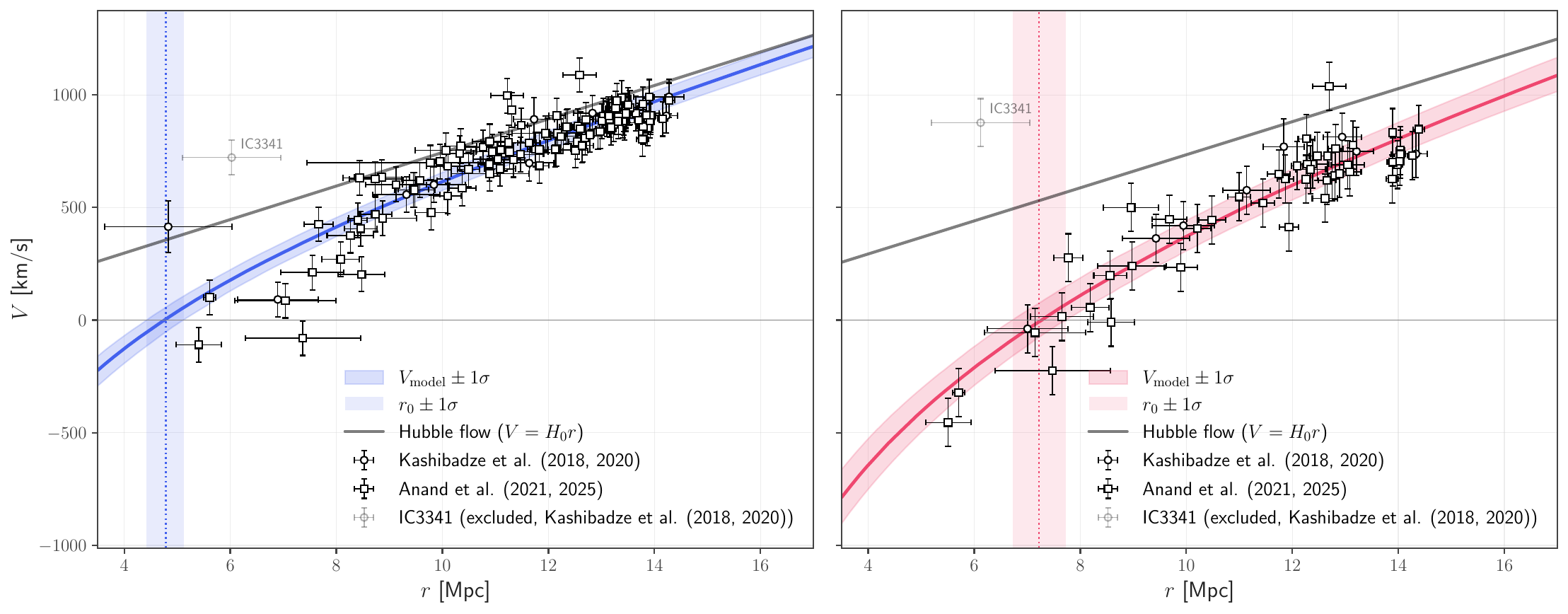}
\caption{Hubble flow fits with five parameters around the Virgo cluster from Tables~\ref{tab:virgo_Anand} and \ref{tab:virgo_Kashibadze} for the \cite{bib:Baushev2020} model using the minor (left) and the major infall model with $| \cos \lambda | > 0.707$ (right). 
The fit parameters are shown in Table~\ref{tab:virgo_H0_comparison_full}. The zero-velocity radius and its 1-$\sigma$ uncertainty are marked by a dotted vertical line and a shaded vertical area. The Hubble flow fit of the combined dataset and its 1-$\sigma$ uncertainty band using the parameters of Table~\ref{tab:virgo_H0_comparison_full} are marked by the solid lines and darker shaded areas around them. Galaxies present only in the ``Kashibadze" sample are marked by circles, all others by squares.}
\label{fig:virgo_data}
\end{figure*}

\begin{figure}
\includegraphics[width=\linewidth]{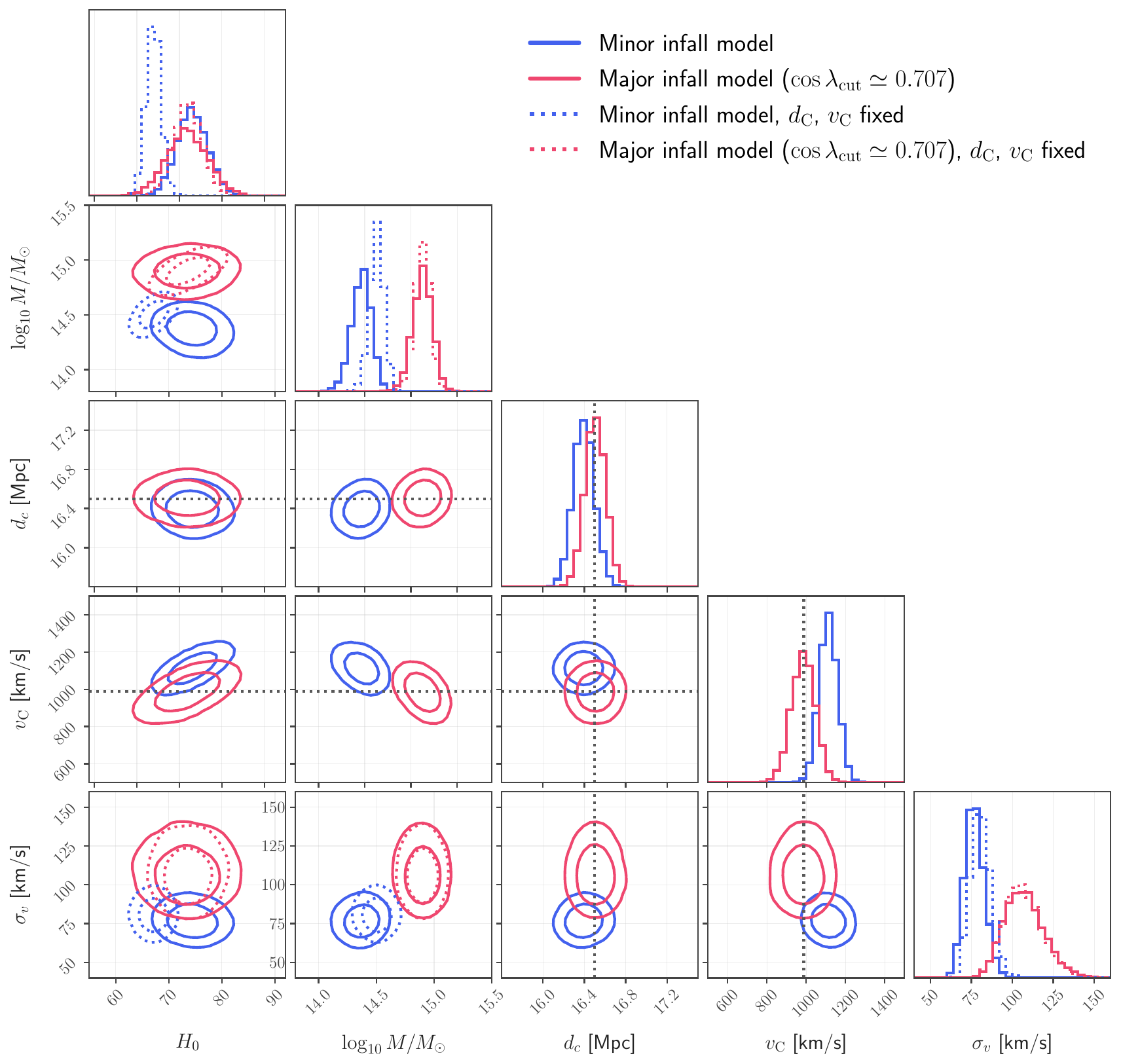}
\caption{Posterior of the Hubble flow fit (prior to bias correction) for the combined dataset. Contours mark the 68\% and 95\% uncertainty bounds of $H_0$, $\log_{10} (M/M_\odot)$, $d_\mathrm{C}$, $v_\mathrm{C}$ and $\sigma_v$. Red contours show the major, blue contours the minor infall model. Solid lines show result when $d_\mathrm{C}$ and $v_\mathrm{C}$ are fitted using a prior and dotted lines show result when $d_\mathrm{C}$ and $v_\mathrm{C}$ are fixed. Dotted black line indicates the $d_\mathrm{C}$ and $v_\mathrm{C}$ prior mean values.}
\label{fig:virgo_posterior}
\end{figure}

We now apply the pipeline set up in this section to the observational data. The fits are performed on a Hubble diagram of velocities by a minor infall model without any cuts and by a major infall model with $| \cos \lambda | > 0.707$, using the \texttt{emcee} MCMC sampler.

We first fit the data in Table~\ref{tab:virgo_Anand}, then the data from Table~\ref{tab:virgo_Kashibadze} and, lastly, the joint dataset. 
The Hubble diagram with the best-fit model lines are shown in Fig.~\ref{fig:virgo_data}.
The fitting results are listed in Table~\ref{tab:virgo_H0_comparison_full}. Fig.~\ref{fig:virgo_posterior} shows the corner plot for the combined dataset.
After a first fit, we noted that the galaxy IC3341 was a 4-$\sigma$ outlier, responsible for numerical instabilities, in particular when fitting the major infall model; it is thus excluded from our final fit.

\begin{table*}
\centering
\caption{\label{tab:virgo_H0_comparison_full}
Hubble flow fit results for the Virgo cluster datasets of Tables~\ref{tab:virgo_Anand} and \ref{tab:virgo_Kashibadze} (column 1),
the infall model used (column 2), the number of galaxies used in the fit (column 3), the inferred $H_0$ (column 4) and its de-biased value, as detailed in Section~\ref{sec:virgo_debiasing} (column 5), the inferred $M$ (column 6), the distance to the cluster barycenter $d_\mathrm{C}$ (column 7), the recession velocity of the barycenter $v_\mathrm{C}$ (column 8) and the scatter velocity parameter $\sigma_v$ (column 9). For diagnostic purposes, $\chi^2_\nu$ is also listed (column 10), as expected, $\chi^2_\nu \approx 1$.}
\smallskip
\renewcommand{\arraystretch}{1.15}
\setlength{\tabcolsep}{4pt}
\resizebox{\textwidth}{!}{%
\begin{tabular}{llcccccccc}
\toprule
Dataset & Infall & $n_\mathrm{gal}$ & $H_0$ & $H_0^{\text{bias corr.}}$ & $\log_{10} M / M_\odot$ & $d_\mathrm{C}$ & $v_\mathrm{C}$ & $\sigma_v$ & \mbox{$\chi_\nu^2$} \\
 & model & & $[\mbox{km}\, \mbox{s}^{-1} \, \mbox{Mpc}^{-1}]$ & $[\mbox{km}\, \mbox{s}^{-1} \, \mbox{Mpc}^{-1}]$ & [dex] & [Mpc] & $[\mbox{km}\, \mbox{s}^{-1}]$ & $[\mbox{km}\, \mbox{s}^{-1}]$ & \\
\midrule
\multicolumn{10}{c}{\textbf{5-parameter Hubble flow fit (``Baseline fit")}} \\
\midrule
\multirow[c]{2}{*}{\cite{bib:Anand2021,bib:Anand2025}} & minor & 107 & $73.3^{+3.0}_{-2.8}$ & $80.5^{+7.8}_{-7.7}$ & $14.47^{+0.08}_{-0.09}$ & $16.44^{+0.10}_{-0.10}$ & $1078.6^{+47.5}_{-46.8}$ & $77.0^{+6.5}_{-5.7}$ & \mbox{1.14} \\
 & major & 47 & $73.8^{+3.9}_{-4.0}$ & $87.7^{+8.8}_{-8.8}$ & $14.92^{+0.08}_{-0.09}$ & $16.50^{+0.10}_{-0.10}$ & $986.0^{+60.1}_{-59.9}$ & $104.8^{+14.0}_{-11.5}$ & \mbox{1.13} \\
\midrule
\multirow[c]{2}{*}{\cite{bib:Kashibadze2018,bib:Kashibadze2020}} & minor & 70 & $72.5^{+3.7}_{-3.6}$ & $79.8^{+8.1}_{-8.1}$ & $14.40^{+0.10}_{-0.11}$ & $16.42^{+0.10}_{-0.10}$ & $1083.4^{+57.8}_{-57.7}$ & $65.2^{+10.0}_{-8.2}$ & \mbox{1.47} \\
 & major & 36 & $73.6^{+4.5}_{-4.3}$ & $87.5^{+9.1}_{-8.9}$ & $14.89^{+0.09}_{-0.10}$ & $16.50^{+0.10}_{-0.10}$ & $990.7^{+59.4}_{-58.5}$ & $79.2^{+15.2}_{-12.5}$ & \mbox{1.26} \\
\midrule
\multirow[c]{2}{*}{Combined} & minor & 121 & $74.4^{+3.0}_{-2.8}$ & $81.6^{+7.8}_{-7.8}$ & $14.37^{+0.08}_{-0.09}$ & $16.40^{+0.10}_{-0.10}$ & $1113.6^{+46.2}_{-45.9}$ & $76.7^{+6.1}_{-5.6}$ & \mbox{1.20} \\
 & major & 55 & $73.4^{+3.9}_{-3.8}$ & $87.3^{+8.8}_{-8.8}$ & $14.90^{+0.08}_{-0.09}$ & $16.51^{+0.10}_{-0.10}$ & $985.8^{+58.0}_{-60.9}$ & $102.2^{+12.9}_{-11.2}$ & \mbox{1.12} \\
\midrule
\multicolumn{10}{c}{\textbf{3-parameter Hubble flow fit}} \\
\midrule
\multirow[c]{2}{*}{\cite{bib:Anand2021,bib:Anand2025}} & minor & 107 & $68.4^{+1.5}_{-1.5}$ & $75.6^{+7.9}_{-7.9}$ & $14.56^{+0.06}_{-0.06}$ & $16.50$ & $988.0$ & $80.4^{+6.6}_{-5.8}$ & \mbox{1.12} \\
 & major & 47 & $73.7^{+3.1}_{-2.9}$ & $86.1^{+9.4}_{-9.4}$ & $14.92^{+0.07}_{-0.07}$ & $16.50$ & $988.0$ & $109.4^{+13.9}_{-12.0}$ & \mbox{1.10} \\
\midrule
\multirow[c]{2}{*}{\cite{bib:Kashibadze2018,bib:Kashibadze2020}} & minor & 70 & $67.1^{+1.5}_{-1.6}$ & $74.3^{+7.9}_{-7.9}$ & $14.50^{+0.07}_{-0.08}$ & $16.50$ & $988.0$ & $70.5^{+10.3}_{-9.0}$ & \mbox{1.44} \\
 & major & 36 & $73.4^{+3.4}_{-3.0}$ & $85.7^{+9.5}_{-9.4}$ & $14.89^{+0.09}_{-0.09}$ & $16.50$ & $988.0$ & $85.4^{+15.2}_{-12.9}$ & \mbox{1.18} \\
\midrule
\multirow[c]{2}{*}{Combined} & minor & 121 & $67.1^{+1.3}_{-1.4}$ & $74.3^{+7.8}_{-7.8}$ & $14.50^{+0.06}_{-0.06}$ & $16.50$ & $988.0$ & $80.5^{+6.6}_{-5.8}$ & \mbox{1.19} \\
 & major & 55 & $73.6^{+2.7}_{-2.7}$ & $86.0^{+9.3}_{-9.3}$ & $14.90^{+0.06}_{-0.07}$ & $16.50$ & $988.0$ & $106.2^{+12.1}_{-10.6}$ & \mbox{1.08} \\
\bottomrule
\end{tabular}}
\end{table*}

\begin{figure}
    \centering
    \includegraphics[width=0.95\columnwidth]{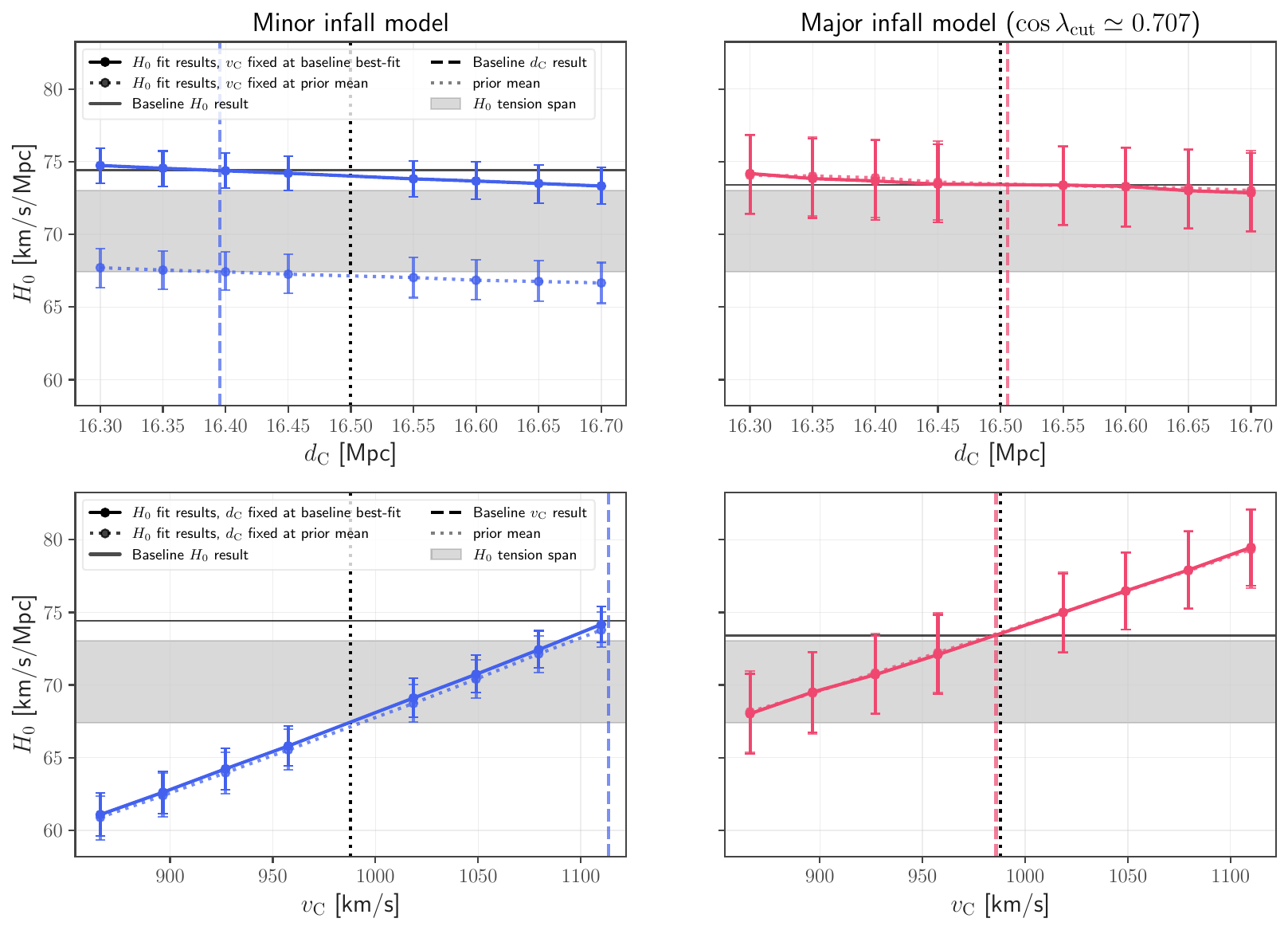}
    \caption{$H_0$ dependence on $d_\mathrm{C}$ (top panel) and $v_\mathrm{C}$ (bottom panel) variations for the major (right panel, red) and minor (left panel, blue) infall models. In these fits $d_\mathrm{C}$ and $v_\mathrm{C}$ are not fitted but fixed either to the baseline result (solid line) or prior mean (dotted line) if not varied. The variation corresponds to $\pm$ 0.5, 1, 1.5, 2$\sigma$ of the priors of $d_\mathrm{C}$ and $v_\mathrm{C}$.}
    \label{fig:dc_vc_sys}
    \vspace{2ex}
    \includegraphics[width=0.8\columnwidth]{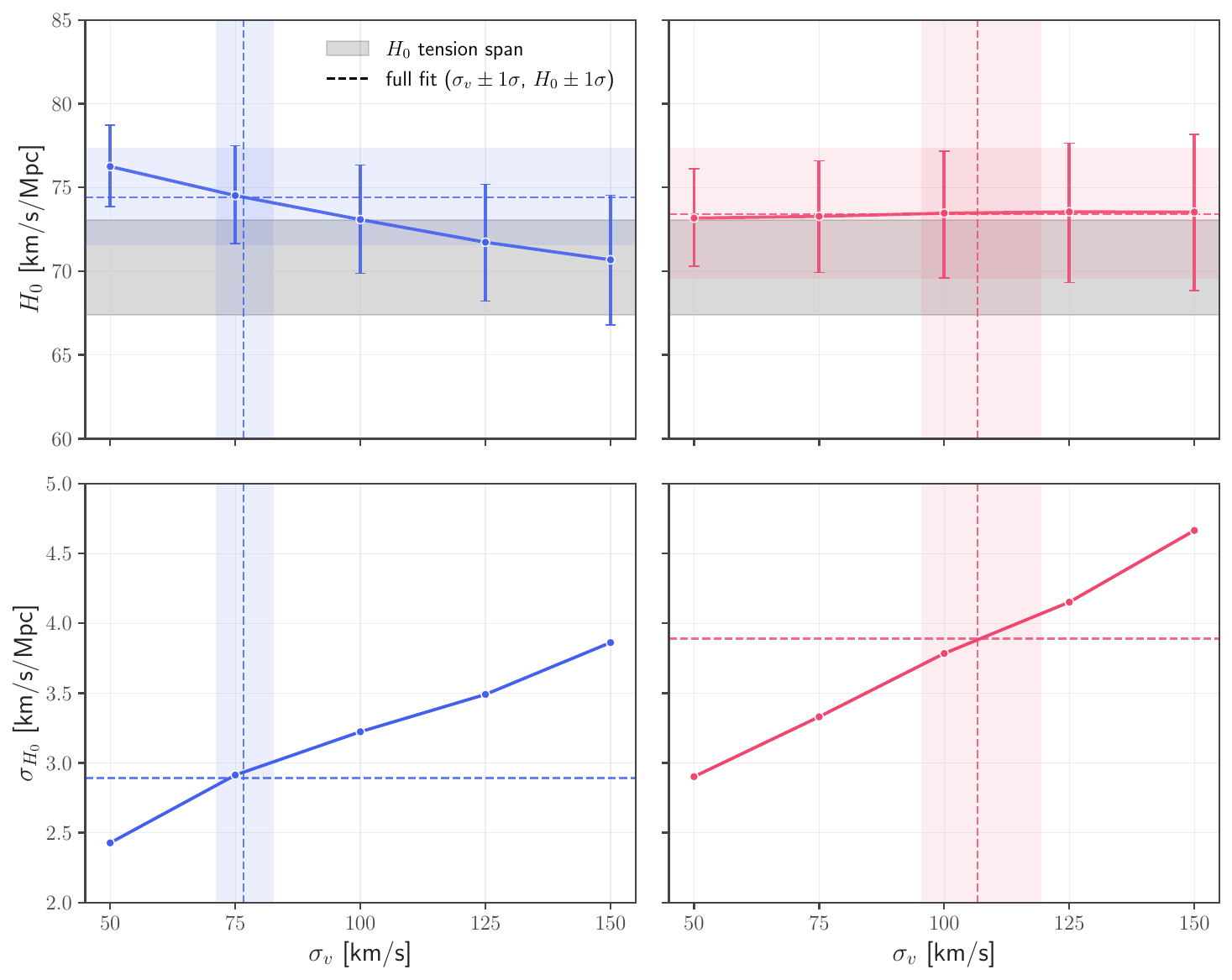}
    \caption{$H_0$ dependence on $\sigma_v$ for the minor infall (blue) and the major infall (red). Baseline results for $H_0$ and $\sigma_v$ are represented by dotted lines.}
    \label{fig:h0_sigmav}
\end{figure}

Irrespective of taking the 3- or 5-parameter fit, the two datasets have $H_0$-values that agree within their uncertainty bounds for each infall model, so that a combined dataset and fit is done with the largest number of TRGB-based distances used. 
Comparing the size of the error bars for the three datasets, we note that an increased number of satellite galaxies does not significantly improve the uncertainties anymore.
This implies that further increasing the amount of satellites will not put tighter constraints on $H_0$. 

To further inspect the source of error on the $H_0$ measurement,  we computed a $H_0$ error budget decomposition for our fit of the minor and the major infall. To obtain each component $k$ we re-run the fit by setting the error component to zero and computing: $\sigma_k^2 = \sigma^2_\text{all}- \sigma_{\text{without } k}^2$. For $d_\mathrm{C}$ (resp. $v_\mathrm{C}$) we follow the same methodology by fixing $d_\mathrm{C}$ (resp. $v_\mathrm{C}$) to the best-fit value and running the fit. We found that, in the 5-parameter fit, the main components of the error were the unknown velocity scatter $\sigma_v$ and the determination of Virgo's co-moving frame $v_\mathrm{C}$, each accounting for almost $50\%$ of the variance on $H_0$.  The uncertainty on distances and velocities of galaxies is negligible, accounting for less than 1\% of the variance. The same budget for the 3-parameter fit leads to an error budget only dominated by $\sigma_v$.

A comparison of the 5- and 3-parameter fits shows similar results in the case of the major infall while the minor infall model has a shift of $\sim 7$ km\,s$^{-1}$\,Mpc$^{-1}$ on $H_0$ between the two fits.
This can be explained by the existing degeneracy between $H_0$ and $v_\mathrm{C}$ as pointed out by the $H_0 - v_\mathrm{C}$ corner plot in Fig.~\ref{fig:virgo_posterior}. For the 5-parameter fit, while the posterior of $d_\mathrm{C}$ falls back on the prior for both the minor and major infall models, the posterior of $v_\mathrm{C}$ is compatible with the prior only for the major infall and we observe a shift of $v_\mathrm{C}$ from its prior value of about $+2\sigma$ for the minor infall.
Applying a $\cos\lambda_{\rm cut}\simeq 0.707$-cut to the 5-parameter fit for the minor infall model and thereby restricting it more strongly to Virgo's line-of-sight, we find $v_\mathrm{C} = 994^{+61}_{-63}$\, km/s, compatible with the prior. 
This result points towards a possible bias of the minor infall model that we were not able to detect in our simulated mocks.

 The $H_0$--$v_\mathrm{C}$ degeneracy can be better quantified with Fig.~\ref{fig:dc_vc_sys}, which shows the result of $H_0$ when $d_\mathrm{C}$ and $v_\mathrm{C}$ are fixed while other parameters are fitted. 
 We see in the top panel that the dependence of $H_0$ on $d_\mathrm{C}$ is not very strong, while the bottom panel shows that $H_0$ is strongly correlated to $v_\mathrm{C}$. 
 In particular, we see that varying $v_\mathrm{C}$ in a range of $\pm1\sigma$ around the prior mean shifts $H_0$ by $\pm 3 \, \mbox{km}\,\mbox{s}^{-1}\, \mbox{Mpc}^{-1}$. This last result shows that the error from the 3-parameter fit by neglecting the $H_0$--$v_\mathrm{C}$ correlation, underestimates the error on $H_0$. We conclude that angular bias in the minor infall model needs to be further investigated in future works and that tighter constraints on $v_\mathrm{C}$ are required to improve the constraints on $H_0$.

In Fig.~\ref{fig:h0_sigmav} we show the dependence of the fitted $H_0$ in the case where $\sigma_v$ is fixed in a range from $50$--$200\,\mbox{km}\,\mbox{s}^{-1}$. 
For the minor infall model, we see that $H_0$ is decreasing for a higher $\sigma_v$-value, while for the major infall, $H_0$ remains independent of $\sigma_v$ values and only its uncertainties grow for increasing $\sigma_v$. 
Results show that fixing $\sigma_v$ to a value close to the fitted value gives consistent results; moreover, in the considered range, all $H_0$ values are within 1$\sigma$ of our baseline result. 
Fig.~\ref{fig:h0_sigmav} also shows that $\sigma_{H_0}$ is increasing with $\sigma_v$ for both models, as expected.

\begin{figure}
\centering
\includegraphics[width=\columnwidth]{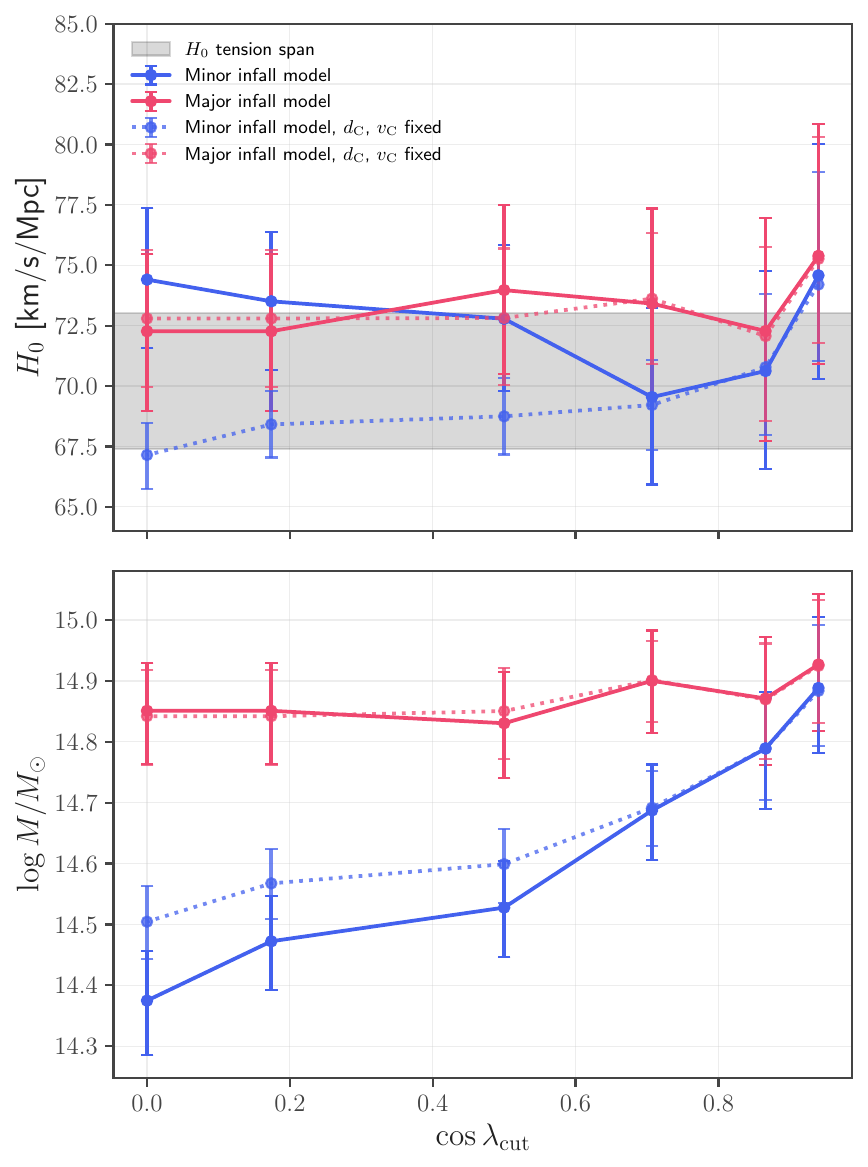}
\caption{\label{fig:virgo_coscut}Impact of a $\lambda$-cut on the inferred $H_0$ (top) and cluster mass (bottom) for the combined dataset of the Virgo cluster. 
Using the minor infall model in the Hubble flow fit is marked in blue, using the major infall model in red. The gray band in the top panel indicates the range spanned by current early- and late-Universe $H_0$-values. Solid lines show result when $d_\mathrm{C}$ and $v_\mathrm{C}$ are fitted using a prior and dotted lines show result when $d_\mathrm{C}$ and $v_\mathrm{C}$ are fixed.}
\end{figure}

In Fig.~\ref{fig:virgo_coscut} we show the dependence of $H_0$ on the $\cos\lambda$ cut. 
We can see that the major model results do not show a strong dependence on the $\cos\lambda_\text{cut}$. 
For the minor infall, we observe small variations of $H_0$, within $\sim 1 \sigma$ from the baseline results across the $\cos\lambda_\text{cut}$ values. However, looking at $M$, the major infall model yields stable results and the minor infall model shows a strong dependence of $\log M / M_\odot$ with respect to $\lambda_\text{cut}$. 
We note that the inferred parameters coincide within their uncertainty bounds for $\cos\lambda_\text{cut} \rightarrow 1$ as it is expected because both infall models converge in that limit. 

\subsection{De-biasing}
\label{sec:virgo_debiasing}

Assuming that the observations of the Virgo cluster represent a typical dataset in the mock observations of Section~\ref{sec:virgo_mock}, we can use the simulated data to de-bias the $H_0$-value of the observations. For the sake of simplicity, we assume that the bias is additive. The study of simulations with varying $H_0$ to test this assumption is left for future work.

We estimate the bias $b_{H_0}$ and the halo population scatter $\sigma_{\text{halo}}$ from Fig.~\ref{fig:mock_lambda_cuts} by maximizing the Gaussian likelihood $\prod_h \mathcal{L}_h$ such that for each halo $h$,
\begin{equation}
    \mathcal{L}_h = \mathcal{N}\left(\Delta H_{0, h} - b_{H_0}, \sqrt{\sigma_{H_0, h}^2 + \sigma_{\text{halo}}^2}\right) \;.
\end{equation}
The halo population scatter in $H_0$, $\sigma_{\text{halo}}$, encodes the unmodeled halo-to-halo variance in the mock data. 
We attribute the origin of this variance to the variety of cluster morphologies and evolution histories.  
Further investigations are necessary to determine if a better selection of the class of clusters in which the model can be used or if a generalization of the spherical collapse model can reduce $\sigma_{\text{halo}}$.

Given the biases and halo population scatter for the minor and major infall models as stated in Section~\ref{sec:virgo_mock}, we subtract the average bias from the $H_0$-values inferred using the two infall models to obtain the $H_0^\mathrm{bias \, corr.}$-values in Table~\ref{tab:virgo_H0_comparison_full}. 
As already mentioned in Section~\ref{sec:virgo_mock}, the de-biasing of a single structure like Virgo requires a better characterization of the representative halo population and a study that shows the dependence of the bias on quality cuts. Yet, it is a first attempt to develop a de-biasing approach.

\begin{table*}
\caption{Earlier $H_0$ inferences for the Virgo cluster using Eq.~\eqref{eq:Peirani} for the Hubble flow fit. $r_\mathrm{min}$ and $r_\mathrm{max}$ denote the distances from the Virgo center where the fit begins and ends, respectively. 
$n_\mathrm{gal}$ is the number of galaxies used in the fit.}

\centering
\label{tab:prior_H0}
\begin{tabular}{lccccccccc}
\hline
Reference & Distance & $\left[r_\mathrm{min}, r_\mathrm{max}\right]$ & $n_\mathrm{gal}$ & \multicolumn{3}{c}{$H_0 \pm (\Delta H_0)_\mathrm{stat} \pm (\Delta H_0)_\mathrm{sys}$} & \multicolumn{3}{c}{Parameters for Eq.~\eqref{eq:Peirani}} \\
 & probe & $[$Mpc$]$ & & \multicolumn{3}{c}{$[$km/s/Mpc$]$}  & A & n & B  \\
\hline
\cite{bib:Peirani2006} & TF & $[1.7,15]$ & 27 & 65 & 9 & -- & $0.875$ & $0.865$ & $1.274$ \\
\cite{bib:Peirani2008} & TF & $[3.5,15]$ & -- & 71 & 9 & -- & $0.976$ & $0.627$ & $1.377$ \\
\cite{bib:Peirani2008} & TF & $[3.5,15]$ & -- & 92 & 12 & -- &  $1.038$ & $0.5$ & $1.196$ \\
\cite{bib:Kim2020} & TRGB & $[5,13]$ &  33 & 65.8 & 3.5 & 2.4 & $1.1$ & $0.5$ & $1.494$ \\
\hline
\vspace{2ex}
\end{tabular}

\end{table*}

\section{Discussion \& Conclusion}
\label{sec:discussion}

The local Hubble flow around a bound structure is not a pure linear expansion.
For galaxies in the infall region, their peculiar velocities, sourced by the collapsing overdensity, decouple from the cosmic recession, producing a characteristic turnover of the velocity-distance relation. 
Its zero-crossing defines the zero-velocity radius, $r_0$. In this work, we showed that the analytical model of the Hubble flow from $r_0$ into the embedding background, \cite{bib:Baushev2020}, can be used to infer $H_0$ and the mass of the collapsed structure $M$ (or its zero-velocity radius, $r_0$) in Hubble flow analyses. 
We also showed that it should be preferred over the computationally cheaper, approximating solutions like the ones from \cite{bib:Peirani2006}, \cite{bib:Peirani2008} or \cite{bib:Penarrubia2014} as it yields a more accurate reconstruction, particularly in the transition to the embedding background.  
As derived in this work, the fitting functions can be described in one formula, Eq.~\eqref{eq:Peirani}, that contains up to three heuristically inferred parameters.
Even though \cite{bib:Penarrubia2014} employ different prerequisites than \cite{bib:Peirani2006} and \cite{bib:Peirani2008}, they arrive at the same approximation and only choose different parameter values. 

Since the accuracy and precision of the inferred $H_0$ and $M$ from such Hubble flow fits have never been systematically analysed, this work uses the IllustrisTNG300 simulation to show the bias and halo-to-halo variation of these fitting functions in a realistic setting. Dependencies of the bias and the variation on the range of distances, the number of galaxies, and the halo masses used in the Hubble flow fit are also investigated to clearly outline the limitations of the spherical collapse models with their additional assumptions.

The theoretical derivations of Section~\ref{sec:models} already show that the widely used analytic approximations of \citet{bib:Peirani2006, bib:Peirani2008} and \citet{bib:Penarrubia2014} do not reconstruct the unperturbed cosmological Hubble flow at large distances from a spherically collapsing structure.  
Their fitting function, Eq.~\eqref{eq:Peirani}, has an asymptote towards $v\to B\,H_0 r$ with $B=1.377$ and $B=1.494$, i.e.~$38\%$ and $49\%$ above $H_0 r$, respectively, see Fig.~\ref{fig:hubble_flow_stacked}. 
Since the fit anchors $H_0$ to this asymptote, an overestimated $B$ biases the inferred Hubble constant low. 
This result is independently supported by the simulation-calibrated analysis of \citet{bib:Benisty2026a} which fitted simulated data from TNG-halos to Eq.~\eqref{eq:Peirani}, solving for the coefficients $A$, $B$, and $n$ with given $H_0$ and $M$.
They obtained $A\simeq B\simeq1$ with $n$ unconstrained by the halo statistics. 
Thus, the data themselves favored the asymptotic slope $B=1$ that the exact solution satisfies identically.

In this work, for the first time, we implement a Hubble flow fit directly based on the spherical collapse integration of \citet{bib:Baushev2020}, Eq.~\eqref{eq:Baushev}.
Its asymptote is $v\to H_0 r$ by construction, so that this Hubble flow fit models the cosmological background Hubble flow within the observed range. 
While this model is computationally more intensive, it does not introduce an asymptotic bias and therefore does not require any heuristically inferred parameters in the fit. 

In Section~\ref{sec:simulation}, we apply the \cite{bib:Baushev2020} solution and the \cite{bib:Penarrubia2014} approximation to Hubble flow fits of Illustris TNG300  simulated isolated halos having masses between $10^{11}~M_\odot < M_{200} < 10^{15}\,M_\odot$. 
In this ideal setting with known $H_0$, $M_{200}$, and the full three-dimensional velocity information of all infalling galaxies, we show that $H_0$-values inferred from Eq.~\eqref{eq:Peirani} suffer from a significant negative bias. 
While this could have been expected, we also find the $H_0$-values from the \cite{bib:Baushev2020} model to be biased.
The bias is smaller than for the \cite{bib:Penarrubia2014} model, yet, still significant, see Fig.~\ref{fig:H0_bias}. 
A further hint that corroborates the spherical collapse model
to be an oversimplified model is the fact that the average uncertainty on $H_0$ for an individual fit is much smaller than the standard deviation obtained from all inferred $H_0$ of the TNG300 halo population. 
In order to extract $H_0$ from observables of Hubble flows, we thus conclude that a de-biasing is necessary if we base our cosmology on $\Lambda$CDM and consider the simulation test done in this work as a representative sample.

Having determined the limits of the Hubble flow fitting model of \cite{bib:Baushev2020} as an $H_0$-estimator in the ideal setting, we apply it to the most extensive dataset of 121 precise TRGB-based distances and line-of-sight velocities of galaxies falling into the Virgo cluster. 
Appendix~\ref{app:data_virgo} provides our datasets based on \cite{bib:Kashibadze2018,bib:Kashibadze2020,bib:Anand2021} and \cite{bib:Anand2025}. 
Those two samples have different calibrations, yet, as we show, they can still be combined as their $H_0$-values agree within the error bounds. 
To compensate the absence of 3D data to describe galaxy motions, we use the minor and major infall models to convert the line-of-sight velocities into the Virgo's reference frame~\citep{bib:Karachentsev2006,bib:Wagner2025}.
These models only maximally use the observables when the infall angles $\lambda$ fulfill $|\cos \lambda | \approx 1$, which is why we also investigate the impact of galaxies at higher infall angles in mock observations. 

We select halos with $M_{200}>10^{14}\,M_\odot$ from the Illustris TNG300 simulation as Virgo mock data and place them to mimic our observers' location towards Virgo.
We considered the error on the position and velocity of the main halo by including the $d_\mathrm{C}$ and $v_\mathrm{C}$ as free parameters of the fit constrained using informative priors from previous measurements. We compared this approach with the usual 3-parameter fit that fixes those two parameters.
We find that the bias on $H_0$ is minimal if the minor infall model is used without any cuts on $\lambda$. In contrast, the major infall model yields a larger bias, suffers from a larger halo-to-halo variation, and is highly sensitive to $\lambda$-cuts.

We applied the same methodology to Virgo's galaxy data and obtained the results summarized in Table~\ref{tab:virgo_H0_comparison_full}.   We showed that the main source of statistical uncertainty is the residual scatter to the Hubble flow model while systematic uncertainties are dominated by the degeneracy between the velocity of the Virgo cluster and $H_0$. This latter result is supported by the comparison between the results of the 5- and 3-parameters fit.

Even though the minor infall model was found to be numerically more robust with smaller uncertainties on $H_0$ than the major infall model for our simulated mocks, we found that it is highly sensitive to the $H_0$--$v_\mathrm{C}$ degeneracy (see Fig.~\ref{fig:dc_vc_sys}).
This is why the $H_0$-values of the 3-parameter fit obtained a lower $H_0=67.1^{+1.3}_{-1.4}\, \mbox{km}\, \mbox{s}^{-1} \, \mbox{Mpc}^{-1}$ compared to the 5-parameter fit with $H_0=74.4^{+3.0}_{-2.8}\, \mbox{km}\, \mbox{s}^{-1} \, \mbox{Mpc}^{-1}$, which increased $v_\mathrm{C}$.
Thus, neither of the two infall models can be favored over the other, but rather the joint information from both is required to fully capture all biases in the $H_0$-inference.

Taking the simulated mock to be representative of the Virgo cluster, we also presented a first approach to de-bias the $H_0$-value in Section~\ref{sec:virgo_debiasing}.  
Whether this bias correction yields a more accurate estimate of $H_0$ remains a question of the mock data being a representative of the structure for which we want to de-bias $H_0$. 
Since Virgo is only a single, most likely not typical structure to apply the \cite{bib:Baushev2020} model, further studies are required with larger samples of observed clusters and refined simulations.

Comparing our results to existing ones is not straightforward because of the different selections one may make from the dataset, including its calibration and transformation into the co-moving reference frame, to the implementation of the Hubble flow fit. 
This could be the reason why there are only a few related works that constrained $H_0$ from the Virgo cluster, as shown in Table~\ref{tab:prior_H0} based on \cite{bib:Peirani2006, bib:Peirani2008}, and \cite{bib:Kim2020}.

Although a direct comparison to these prior works is not possible, our work gives a deeper insight into the physical reasons behind the trends observed in all three works and also in \cite{bib:Karachentsev2018}: our results in the 3-parameter fit together with Fig.~\ref{fig:dc_vc_sys} explain that the $H_0$-value obtained using the minor infall model is lower than the one using the major infall model due to the fixed $v_\mathrm{C}$. Leaving $v_\mathrm{C}$ open to be fitted alongside $H_0$ and removing the $\lambda$-cut, the $H_0$--$v_\mathrm{C}$ correlation might break this order. Beyond the Virgo-example, this understanding is equally valid for the results gained from other structures, e.g., in \cite{bib:Wagner2026} for the M~81-group and \cite{bib:Faucher2026} for the Centaurus~A group.

Moreover, we showed that $\lambda$-cuts, as particularly employed in \cite{bib:Kim2020}, can make the inferred $H_0$-values of the major and minor infall models converge to a common value. 
Yet, when tested with simulations, the $H_0$ values obtained for different $\lambda$-cuts are still not equal to $H_0^{\mathrm{true}}$, as shown in Fig.~\ref{fig:lambdacut}. The resulting $H_0$ is rather subject to a larger average bias compared to applying smaller or no cuts. 
We strictly prove this trend only for the \cite{bib:Baushev2020} model, however, the \cite{bib:Penarrubia2014} approximation used in \cite{bib:Kim2020} showed similar behavior in the $H_0$-bias over all tested halo masses in Section~\ref{sec:simulation}.

To improve the uncertainty, development of more specific simulations for this measurement is required to better characterize the bias of halos with similar characteristics as the cluster used for the measurement. A more general collapse model would also help reduce the uncertainty. Spectroscopic surveys across multiple wavebands such as JWST, DESI, and 4MOST provide thousands of line-of-sight velocities per cluster as well as currently most precise TRGB- or SBF-based distances. The status quo of such programs is summarized in \cite{bib:Casertano2026} and studies of Fornax and Virgo are already being pursued \citep{bib:Anand2024,bib:Anand2025} and planned for Coma \citep{bib:Jensen2024}. 
Reaching a high completeness in the volume with 300~Mpc radius around our location, we will be able to test the predictions of $\Lambda$CDM and the entire class of homogeneous and isotropic background cosmologies instead of calibrating potential biases on their simulations.

\begin{acknowledgments}
We thank Gagandeep Anand, Thomas Buchert, Igor D. Karachentsev and Louis E. Strigari for useful discussions and suggestions. D.B.~is supported by the COST Action CA24101. J.W.~gratefully acknowledges the support of an O'Shaughnessy Ventures Fellowship. 
\end{acknowledgments}

\software{
    iminuit \citep{bib:Dembinski2020},
    Minuit \citep{bib:James1975},
    emcee \citep{bib:Foreman},
    NumPy \citep{bib:Harris2020},
    SciPy \citep{bib:Virtanen2020},
    Astropy \citep{2013A&A...558A..33A, 2018AJ....156..123A, bib:Astropy2022},
    Matplotlib \citep{bib:Hunter2007},
    corner \citep{bib:ForemanMackey2016}
}

\bibliography{ref.bib}
\bibliographystyle{aasjournal}

\appendix
\section{Derivation of the Hubble flow v(r)-relation}
\label{app:derivation_HFF}
In the following, we use the spherically symmetric Lemaître-Tolman-Bondi space-time metric \citep{bib:Lemaitre1933, bib:Tolman1934, bib:Bondi1947} for pressure-less dust and a cosmological constant, $\Lambda = 3 \Omega_\Lambda \left(H_0/c\right)^2$, as the basis for our derivations:
\begin{equation}
\mathrm{d}s^2 = - c^2 \mathrm{d}t^2 + \frac{r'(R,t)^2}{1+2E(R)/c^2} \mathrm{d}R^2 + r^2(R,t) \mathrm{d}\Omega^2 \;.
\label{eq:LTB_metric}
\end{equation}
The metric results from solving the Einstein equations in a coordinate frame co-moving with the matter, such that $R$ is the co-moving radius and $r$ the physical radius.
We impose the standard synchronous gauge, which is possible because all dust particles move on geodesics without acceleration.
Primes denote partial derivatives with respect to the co-moving radius, $R$, dots partial derivatives with respect to the cosmic time, $t$. 
As usual, when constraining the coefficient of $\mathrm{d}R^2$, we set the free function of $R$ in the denominator equal to $1+2E(R)/c^2$, such that the radial Einstein equation amounts to the energy conservation
\begin{equation}
\dot{r}^2 = \frac{2GM(R)}{r} + \frac{\Lambda}{3} c^2 r^2 + 2 E(R) 
\label{eq:energy_conservation}
\end{equation}
with the mass density being
\begin{equation}
\rho(R,t) = \frac{1}{4\pi r^2} \frac{M'(R)}{r'} \;.
\label{eq:LTB_density}
\end{equation}
Eq.~\eqref{eq:energy_conservation} is equivalent to Eq.~(2) of \cite{bib:Baushev2020}, when we set $2E(R)$ equal to the last term containing the energy contribution of constant curvature in Eq.~(2) of \cite{bib:Baushev2020}. 
In contrast to the general LTB metric, \cite{bib:Baushev2020} assumes that the matter is moving in separate shells without crossing, i.e.~there is no point in the metric at which $r'(R,t) = 0$ when $M'(R) \ne 0$.  
Co-moving with such a mass shell, $\dot M = 0$, differentiating Eq.~\eqref{eq:energy_conservation} with respect to time once more yields the second-order equation of motion:
\begin{equation}
\ddot{r} = - \frac{GM(R)}{r^2} + \frac{1}{3} \Lambda c^2 r \;.
\label{eq:second_order}
\end{equation}
This is Eq.~(1) of \cite{bib:Peirani2008} and integrating once, their Eq.~(2) is obtained which is equivalent to Eq.~\eqref{eq:energy_conservation} above.
Reducing the equations of \cite{bib:Peirani2006} to a constant dark energy, i.e.~$\Lambda$, this work agrees with \cite{bib:Peirani2008}. 
Thus, under all these prerequisites, \cite{bib:Baushev2020} and \cite{bib:Peirani2008} coincide in their description of the spherical collapse and model a subclass of LTB-solutions with the slight difference that the interpretation of the $E(R)$ explicitly incorporates a constant curvature contribution \citep{bib:Baushev2020} or is simply denoted as the total energy of a mass shell \citep{bib:Peirani2008}. 
Both works also determine the mass enclosed within the volume of the zero-velocity radius, $M(r_0)$, by the same equations. 

The biggest difference between \cite{bib:Baushev2020} and \cite{bib:Peirani2008} is their approach to set up the Hubble flow fit from this collapse model:
\cite{bib:Baushev2020} assumes all shells to be generated at the cosmic origin $t_\mathrm{ini}=0$ and collapses them individually, mimicking a Friedmann model for each shell. 
This is why dimensionless variables are obtained by scaling with each shell parameters individually.
For each shell, the integrals in Eqs.~(10) and (14) of \cite{bib:Baushev2020} are solved to obtain $v(r)$, so that the Hubble flow is exactly plotted for a structure obeying all requirements mentioned above. 

In contrast, \cite{bib:Peirani2008} use the zero-velocity radius, $r_0$, as the reference and scaling radius for the dimensionless equations and take the initial conditions from the matter-dominated era, $r_\mathrm{ini}/r_0 = 0.01$. 
There, Eq.~\eqref{eq:second_order} can be solved analytically to obtain the cosmic time, $t_\mathrm{ini}$.
Their parameter\footnote{not to be confused with our definition of $A$ as the heuristic fitting coefficient in Eq.~\eqref{eq:Peirani}} $A \equiv 2GM/(H_0^2 r_0^3)$ in the dimensionless versions of Eqs.~\eqref{eq:energy_conservation} and \eqref{eq:second_order} can then be determined under the condition that the left-hand side of Eq.~\eqref{eq:energy_conservation} is zero for $r=r_0$ at $t=t_0$ today and that the formation time of the structure's current collapse status is equal to the age of the embedding. 
Thus, $M$ in \cite{bib:Peirani2008} clearly refers to $M(r_0)$.
Subsequently, based on the numerical solutions of Eq.~\eqref{eq:second_order}, they set up a phenomenological fitting function
\begin{equation}
v(r) = - \frac{0.976 H_0}{r^n} \left(\frac{GM}{H_0^2} \right)^{(n+1)/3} + 1.377 \, H_0 r \;,
\label{eq:fitting_function}
\end{equation}
which fulfills $v(r_0)=0$ numerically and allows to constrain $M$ and $H_0$ in a fit without solving computationally-intensive integrals for each shell.
The numerical coefficients incorporate $A=3.7683$ for $\Omega_\Lambda=0.7$ and the exponent $n=0.627$ as an empirical choice. 
While \cite{bib:Peirani2006} and \cite{bib:Peirani2008} are brief in details about the way to solve the differential equation numerically, \cite{bib:DelPopolo2020} and \cite{bib:DelPopolo2022} elaborate on these points and also extend the equations with angular momentum and dynamical friction terms. 
The angular-momentum term is often omitted because it is assumed that group- and cluster-scale structures are rather pressure-supported and the angular momenta of their constituents average out, see, for instance, the discussion in \cite{bib:Song2018} for the limits of this assumption. Assuming spherical symmetry is conserved, adding the angular-momentum term $l^2/r^3$ to the right-hand side of Eq.~\eqref{eq:second_order} would mostly affect the motion close to $M$ as it scales with $r^{-3}$, see \cite{bib:Benisty2024}.

The expanding pull is also observed when calculating the mass enclosed in the zero-velocity radius from 
\begin{equation}
M = M(r_0) = \frac{A H_0^2}{2G} r_0^3 \;,
\end{equation}
yielding larger masses compared to the ones obtained by \cite{bib:Sandage1986} setting $\Lambda=0$, $M = \pi^2 r_0^3/(8G t_0^2)$.
See also \cite{bib:Karachentsev2009, bib:Pavlidou2005,bib:Pavlidou2014} and \cite{bib:Benisty2025} for further usages of these relations.

A similar phenomenological fitting function as Eq.~\eqref{eq:fitting_function} is obtained by \cite{bib:Penarrubia2014}, yet, from a completely different viewpoint.
They use the linear relation 
\begin{equation}
n \approx \sqrt{\frac{GM}{r^3}} t_0 + 0.9 \left(\frac{v}{r} \right) t_0 
\label{eq:linear_Kepler}
\end{equation}
found for Keplerian orbits of physical radius $r$ and velocity $v$ around a point mass $M$ by \cite{bib:Lynden1981} and extend $n = n(\Omega_\Lambda) \approx 1.1 + 0.21 \Omega_\Lambda$ numerically. Inserting it into Eq.~\eqref{eq:linear_Kepler}, the analogue of Eq.~\eqref{eq:fitting_function} reads:
\begin{equation}
v(r) = - 1.1\left(\frac{GM}{r} \right)^{1/2} + 1.494 \, H_0 r \;.
\end{equation}
\cite{bib:Penarrubia2017} rewrite this equation in terms of the zero-velocity radius, also stating that it is not strongly dependent on the impact of the background expansion. 
Subsequently, this leads \cite{bib:Karachentsev2018} to set up another fitting function as 
\begin{equation}
v(r) = - H_\mathrm{loc} r_0 \sqrt{\frac{r_0}{r}} + H_\mathrm{loc} r \;,
\label{eq:karachentsev}
\end{equation}
in which they fit for $r_0$ and $H_\mathrm{loc}$.
The latter is not the Hubble constant of the background cosmology, but an expansion coefficient incorporating local perturbations, which is why we only mention this approach briefly here.

\section{Details of the Hubble flow fitting}
\label{app:implementation}

The Hubble flow fits, particularly the \cite{bib:Baushev2020} integrals, are computationally intensive. 
Therefore, we use the \texttt{Minuit} algorithm as implemented in the Python library \texttt{iminuit}~\citep{bib:James1975,bib:Dembinski2020} for the fitting. 
In order to speed up the fit of \cite{bib:Baushev2020}'s model, we pre-compute the velocity profile on an $N=256^3$ grid of $(10 < H_0 < 150) \, \mbox{km}\,\mbox{s}^{-1}\,\mbox{Mpc}^{-1}$ and $11< \log_{10}(M/M_\odot) < 16$ and $(1.0 < r < 40)$ Mpc.
Pathological fits that are flagged as invalid by \texttt{Minuit} or that are at the boundaries of our parameter intervals are removed.   
To quantify the quality of fit we use the $p$-value computed as the integral over the $\chi^2$-probability: 
\begin{equation}
    p = \int_{\chi^2_{\rm fit}}^{+\infty} \mathcal{P}_{\chi^2}(x; k) dx \;,
\label{eq:p-value}
\end{equation}
in which $k$ is the number of degrees of freedom of the fit.

\begin{figure}[h]
    \centering
\includegraphics[width=\textwidth]{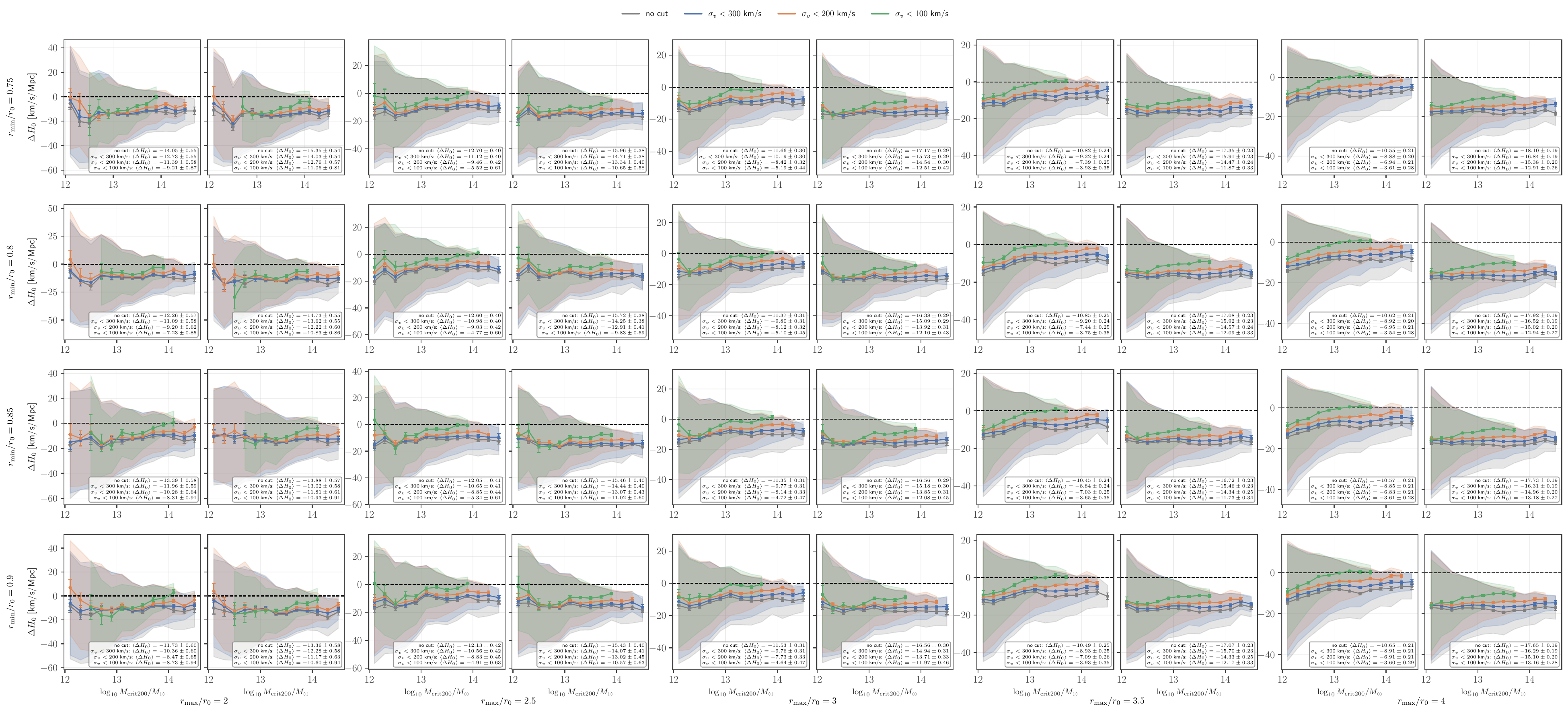}
\caption{Systematic variation of the distance range used for the fit and its impact on the $H_0$-bias for the \cite{bib:Baushev2020} model (first plots) and the \cite{bib:Penarrubia2014} (second plots). The minimum distance for the fits is varied from top to bottom from $0.75\, r_0$ to $0.9\, r_0$, the maximum distance increases from left to right from $2.0\, r_0$ to $4.0\, r_0$.}
    \label{fig:rmin_rmax}
\end{figure}

Running the Hubble flow fits of \cite{bib:Baushev2020} and \cite{bib:Penarrubia2014} for the distance range mentioned in the main text, Fig.~\ref{fig:H0_bias} is obtained. 
Here, we additionally show the impact of the $H_0$-bias as plotted in Fig.~\ref{fig:H0_bias} but for different minimum distances and maximum distances used in the fitting. 
The matrix of plots in Fig.~\ref{fig:rmin_rmax} shows the \cite{bib:Baushev2020} model on the left followed by the \cite{bib:Penarrubia2014} model on the right.

\begin{figure}
   \centering
\includegraphics[width=0.95\linewidth]{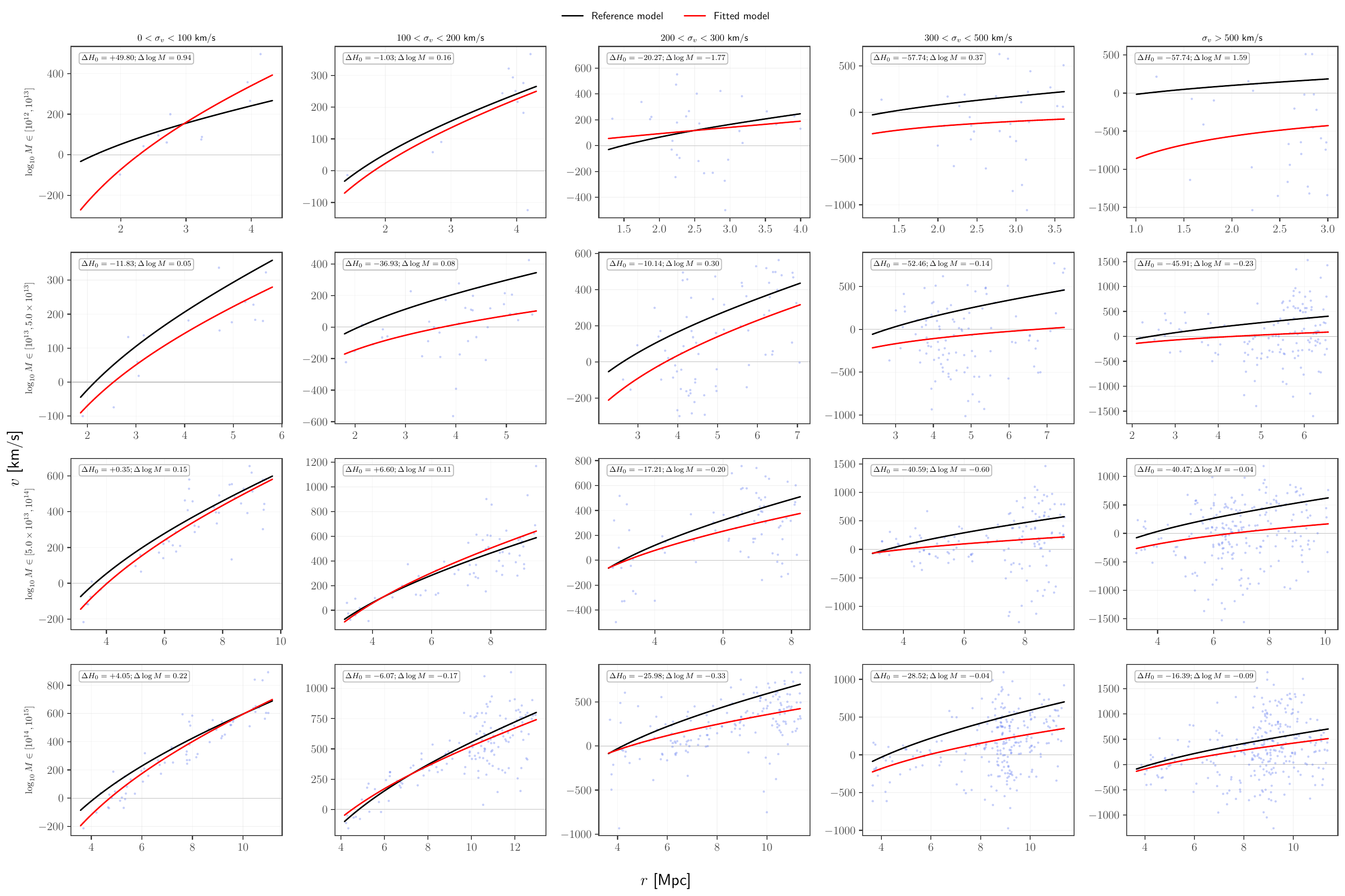}
\caption{
Examples of Hubble flow fit using the Baushev model for different masses and fitted $\sigma_v$. The reference model is obtained using the true value of $H_0$ and $M_{200}$ in the Baushev model (black lines), the fitted model is represented by red lines.}
\label{fig:qof}
\end{figure}

In order to understand which sub-halo configurations are affected by the quality cuts made in Section~\ref{sec:simulation}, we plot a selection of fits according to \cite{bib:Baushev2020} in Fig.~\ref{fig:qof}.
The figure shows that, for instance, for high velocity scatter values, halos have many smaller-scale structures in the Hubble flow that increase its scatter, or have a highly inhomogeneous distribution of sub-halos that deteriorates the fit. 

\section{Virgo Cluster Hubble Flow Data Tables}
\label{app:data_virgo}

\cite{bib:Anand2021} compiled the most recent and joint EDD-TRGB-based distances for 558 galaxies, which was updated in \cite{bib:Anand2025}.
Out of all galaxies, 111 lie within the Virgo field and its Hubble flow.
107 of these galaxies have a distance larger than 4~Mpc from the Virgo barycenter.
Table~\ref{tab:virgo_Anand} shows these galaxies with columns: (1) name, (2) right ascension, (3) declination, (4) heliocentric velocity, (5) measurement uncertainty on the heliocentric velocity, (6) distance from the sun as inferred using the TRGB, (7) measurement uncertainty on the TRGB distance, and (8) apparent integral magnitude of the galaxy in the $B$-band.
 
\cite{bib:Kashibadze2018} collected TRGB-based distances for 80 galaxies.
\cite{bib:Kashibadze2020} updated 17 of these TRGB-based distances. 
From these, only 76 were in the most recent database of \cite{bib:Karachentsev2013} and only 73 also had uncertainties on the measured heliocentric velocities. 
Out of these, 71 have a distance larger than 4~Mpc from the Virgo barycenter. 
Table~\ref{tab:virgo_Kashibadze} shows these galaxies with columns: (1) name, (2) right ascension, (3) declination, (4) heliocentric velocity, (5) measurement uncertainty in the heliocentric velocity, (6) distance from the sun as inferred using the TRGB, (7) measurement uncertainty in the TRGB distance, (8) galaxy type, see \cite{bib:Karachentsev2013} for details, and (9) apparent integral magnitude of the galaxy in the $B$-band.

\startlongtable
\begin{longtable}{lrrrrrrr}
\caption{Dataset of 107 galaxies with TRGB-based distances from \cite{bib:Anand2021} and \cite{bib:Anand2025}, sorted with ascending distance from the Virgo center.\label{tab:virgo_Anand}} \\

\hline
  \multicolumn{1}{c}{Name} &
  \multicolumn{1}{c}{RA} &
  \multicolumn{1}{c}{Dec} &
  \multicolumn{1}{c}{$v_\mathrm{hel}$} &
  \multicolumn{1}{c}{$\sigma_{v}^{\dagger}$} &
  \multicolumn{1}{c}{$d_\mathrm{hel}$} &
  \multicolumn{1}{c}{$\sigma_{d}$} &
  \multicolumn{1}{c}{$B_T$} \\
   \multicolumn{1}{c}{} &
  \multicolumn{1}{c}{$[$deg$]$} &
  \multicolumn{1}{c}{$[$deg$]$} &
  \multicolumn{1}{c}{$[\mbox{km}\, \mbox{s}^{-1}]$} &
  \multicolumn{1}{c}{$[\mbox{km}\, \mbox{s}^{-1}]$} &
  \multicolumn{1}{c}{$[$Mpc$]$} &
  \multicolumn{1}{c}{$[$Mpc$]$} &
  \multicolumn{1}{c}{$[$mag$]$} \\
\hline
\endfirsthead

\multicolumn{8}{c}%
{{\bfseries \tablename\ \thetable{} -- continued from previous page}} \\
\hline
  \multicolumn{1}{c}{Name} &
  \multicolumn{1}{c}{RA} &
  \multicolumn{1}{c}{Dec} &
  \multicolumn{1}{c}{$v_\mathrm{hel}$} &
  \multicolumn{1}{c}{$\sigma_{v}^{\dagger}$} &
  \multicolumn{1}{c}{$d_\mathrm{hel}$} &
  \multicolumn{1}{c}{$\sigma_{d}$} &
  \multicolumn{1}{c}{$B_T$} \\
   \multicolumn{1}{c}{} &
  \multicolumn{1}{c}{$[$deg$]$} &
  \multicolumn{1}{c}{$[$deg$]$} &
  \multicolumn{1}{c}{$[\mbox{km}\, \mbox{s}^{-1}]$} &
  \multicolumn{1}{c}{$[\mbox{km}\, \mbox{s}^{-1}]$} &
  \multicolumn{1}{c}{$[$Mpc$]$} &
  \multicolumn{1}{c}{$[$Mpc$]$} &
  \multicolumn{1}{c}{$[$mag$]$} \\
\hline
\endhead

\hline
\endfoot

\hline \hline
\multicolumn{8}{l}{$^{\dagger}$ The catalogue has no per-object velocity error, so $\sigma_{v} = 3\,\mbox{km}\, \mbox{s}^{-1}$ is adopted for every galaxy.} \endlastfoot

UGC07512 & 186.4212 & 2.1593 & 1506 & 3.0 & 11.59 & 0.43 & 14.61 \\
NGC4565 & 189.0865 & 25.9872 & 1230 & 3.0 & 11.87 & 0.11 & 8.96 \\
PGC043072 & 191.5638 & 10.2055 & 1142 & 3.0 & 9.42 & 0.95 & 15.65 \\
PGC040045 & 185.5317 & 15.7992 & 1298 & 3.0 & 9.08 & 1.09 & 15.56 \\
PGC041395 & 187.7652 & 1.6758 & 1105 & 3.0 & 9.20 & 0.59 & 16.23 \\
NGC4600 & 190.0958 & 3.1177 & 852 & 3.0 & 9.00 & 0.27 & 13.37 \\
NGC4592 & 189.8281 & -0.5320 & 1065 & 3.0 & 8.79 & 0.36 & 11.48 \\
PGC3097708 & 189.3995 & -8.8671 & 1105 & 3.0 & 9.55 & 0.44 & 16.68 \\
PGC044532 & 194.7576 & 34.8608 & 842 & 3.0 & 9.73 & 0.18 & 12.04 \\
PGC042120 & 189.3086 & -10.4979 & 756 & 3.0 & 9.59 & 0.31 & 16.19 \\
NGC4559 & 188.9904 & 27.9597 & 807 & 3.0 & 8.57 & 0.31 & 9.61 \\
NGC4437 & 188.1895 & 0.1146 & 1128 & 3.0 & 8.29 & 0.44 & 9.62 \\
NGC4625 & 190.4697 & 41.2738 & 607 & 3.0 & 10.72 & 0.49 & 12.73 \\
NGC4594 & 189.9977 & -11.6230 & 1024 & 3.0 & 9.35 & 0.31 & 8.70 \\
NGC4627 & 190.4992 & 32.5736 & 542 & 3.0 & 8.60 & 0.52 & 12.93 \\
PGC042115 & 189.3081 & 29.6308 & 758 & 3.0 & 8.24 & 0.65 & 15.90 \\
PGC043851 & 193.4868 & -12.1059 & 824 & 3.0 & 8.95 & 0.58 & 13.64 \\
ESO572-034 & 179.7420 & -19.0299 & 1114 & 3.0 & 10.28 & 0.38 & 13.52 \\
UGC07678 & 188.0010 & 39.8324 & 679 & 3.0 & 8.87 & 0.29 & 14.75 \\
PGC166135 & 185.7197 & 33.8291 & 573 & 3.0 & 7.80 & 0.32 & 17.21 \\
NGC4485 & 187.6303 & 41.7009 & 493 & 3.0 & 8.71 & 2.33 & 11.91 \\
PGC2832120 & 190.8526 & 32.4823 & 735 & 3.0 & 7.41 & 0.31 & 17.53 \\
NGC5055 & 198.9555 & 42.0292 & 484 & 3.0 & 8.83 & 0.08 & 8.89 \\
PGC041093 & 187.2180 & 42.1779 & 407 & 3.0 & 8.32 & 0.57 & 15.07 \\
UGC07827 & 189.9123 & 44.8202 & 553 & 3.0 & 8.87 & 0.41 & \multicolumn{1}{c}{--} \\
NGC4455 & 187.1837 & 22.8215 & 637 & 3.0 & 6.47 & 0.26 & 12.38 \\
PGC041314 & 187.5987 & 42.9015 & 436 & 3.0 & 8.02 & 0.26 & 15.15 \\
PGC166159 & 198.4156 & 42.0436 & 368 & 3.0 & 8.09 & 0.30 & 18.03 \\
PGC086635 & 187.0206 & 22.2908 & 602 & 3.0 & 6.17 & 0.26 & 16.91 \\
NGC4618 & 190.3866 & 41.1489 & 533 & 3.0 & 7.46 & 0.20 & 10.86 \\
SDSSJ121811.07+465500.6 & 184.5461 & 46.9169 & 420 & 3.0 & 8.09 & 0.34 & 16.82 \\
NGC5229 & 203.5120 & 47.9154 & 359 & 3.0 & 8.79 & 0.32 & 13.44 \\
NGC4490 & 187.6516 & 41.6437 & 565 & 3.0 & 6.92 & 0.73 & 8.58 \\
PGC166137 & 186.3678 & 28.4824 & 480 & 3.0 & 5.86 & 0.22 & 16.08 \\
NGC5194 & 202.4695 & 47.1951 & 463 & 3.0 & 8.34 & 0.30 & 8.34 \\
NGC4038 & 180.4709 & -18.8680 & 1642 & 3.0 & 21.05 & 0.73 & 10.36 \\
NGC4242 & 184.3757 & 45.6191 & 506 & 3.0 & 7.31 & 0.33 & 11.26 \\
PGC044681 & 194.9857 & -19.4114 & 825 & 3.0 & 7.11 & 0.29 & 16.47 \\
NGC4258 & 184.7398 & 47.3038 & 448 & 3.0 & 7.54 & 0.09 & 8.47 \\
UGC07408 & 185.3134 & 45.8120 & 463 & 3.0 & 7.11 & 0.07 & 12.81 \\
UGC07903 & 190.9394 & 53.9599 & 443 & 3.0 & 9.46 & 0.52 & 20.10 \\
UGC07356 & 184.7886 & 47.0904 & 132 & 3.0 & 7.11 & 0.07 & 15.82 \\
PGC042134 & 189.3408 & 38.7438 & 458 & 3.0 & 6.00 & 0.22 & 15.26 \\
NGC4248 & 184.4579 & 47.4091 & 484 & 3.0 & 6.88 & 0.43 & 12.44 \\
NGC4144 & 182.4939 & 46.4573 & 262 & 3.0 & 6.73 & 0.12 & 10.76 \\
PGC1726175 & 203.1262 & 25.1236 & 360 & 3.0 & 5.18 & 0.21 & 17.20 \\
PGC044055 & 193.9185 & 19.2092 & 419 & 3.0 & 4.72 & 0.15 & 16.41 \\
NGC5023 & 198.0490 & 44.0381 & 407 & 3.0 & 5.98 & 0.16 & 11.22 \\
PGC044982 & 195.5598 & -17.2378 & 755 & 3.0 & 5.73 & 0.18 & 14.33 \\
PGC2543081 & 179.4759 & 56.6380 & 416 & 3.0 & 8.55 & 0.63 & 17.24 \\
UGC07698 & 188.2266 & 31.5410 & 330 & 3.0 & 4.78 & 0.11 & 12.72 \\
NGC4707 & 192.0956 & 51.1659 & 468 & 3.0 & 6.37 & 0.29 & 13.31 \\
NGC4826 & 194.1825 & 21.6820 & 408 & 3.0 & 4.36 & 0.06 & 8.73 \\
UGC07559 & 186.7707 & 37.1433 & 217 & 3.0 & 4.85 & 0.16 & 13.76 \\
NGC4395 & 186.4537 & 33.5468 & 319 & 3.0 & 4.66 & 0.02 & 9.89 \\
PGC038685 & 182.4847 & 36.4347 & 339 & 3.0 & 4.74 & 0.28 & 15.12 \\
UGC07605 & 187.1613 & 35.7176 & 310 & 3.0 & 4.63 & 0.30 & 14.27 \\
PGC040665 & 186.5709 & 48.4942 & 279 & 3.0 & 5.57 & 0.13 & 14.31 \\
UGC07599 & 187.1171 & 37.2348 & 275 & 3.0 & 4.61 & 0.15 & 14.34 \\
{[KKH2007]J1243+4127} & 190.9819 & 41.4568 & 402 & 3.0 & 4.70 & 0.24 & 16.62 \\
IC3687 & 190.5630 & 38.5026 & 350 & 3.0 & 4.47 & 0.16 & 13.58 \\
UGC08638 & 204.8315 & 24.7752 & 276 & 3.0 & 4.19 & 0.06 & 14.17 \\
ESO443-009 & 193.7232 & -28.3408 & 645 & 3.0 & 5.83 & 0.30 & 16.77 \\
NGC5068 & 199.7295 & -21.0392 & 668 & 3.0 & 5.04 & 0.21 & 10.11 \\
NGC4789A & 193.5225 & 27.1497 & 374 & 3.0 & 3.94 & 0.07 & 14.01 \\
IC4182 & 196.4565 & 37.6050 & 321 & 3.0 & 4.30 & 0.07 & 12.22 \\
NGC4244 & 184.3730 & 37.8072 & 244 & 3.0 & 4.23 & 0.13 & 9.75 \\
NGC4736 & 192.7208 & 41.1200 & 308 & 3.0 & 4.33 & 0.08 & 8.54 \\
PGC042275 & 189.6668 & 32.7669 & 308 & 3.0 & 3.77 & 0.31 & 15.70 \\
PGC166142 & 190.9891 & 43.6616 & 293 & 3.0 & 4.23 & 0.21 & 17.21 \\
UGC08215 & 197.0148 & 46.8280 & 224 & 3.0 & 4.47 & 0.12 & 15.67 \\
NGC4449 & 187.0460 & 44.0930 & 207 & 3.0 & 4.17 & 0.02 & 8.98 \\
NGC3738 & 173.9497 & 54.5251 & 229 & 3.0 & 5.18 & 0.05 & 11.24 \\
PGC166163 & 200.2845 & -31.5291 & 572 & 3.0 & 5.50 & 0.25 & 17.04 \\
ESO444-078 & 204.1284 & -29.2354 & 577 & 3.0 & 5.30 & 0.07 & 14.48 \\
UGC08331 & 198.8740 & 47.4996 & 267 & 3.0 & 4.31 & 0.10 & 13.71 \\
UGC08320 & 198.6174 & 45.9197 & 193 & 3.0 & 4.15 & 0.15 & 11.94 \\
DDO169NW & 198.8337 & 47.5436 & 242 & 3.0 & 4.23 & 0.25 & \multicolumn{1}{c}{--} \\
PGC039228 & 183.9443 & 52.3873 & 159 & 3.0 & 4.51 & 0.19 & 15.04 \\
IC4247 & 201.6852 & -30.3625 & 420 & 3.0 & 5.06 & 0.09 & 13.40 \\
PGC683190 & 173.2957 & -32.9624 & 709 & 3.0 & 5.47 & 0.15 & 16.76 \\
ESO379-007 & 178.6798 & -33.5588 & 644 & 3.0 & 5.32 & 0.17 & 16.15 \\
PGC166170 & 204.1478 & -29.5713 & 390 & 3.0 & 4.90 & 0.18 & 14.11 \\
NGC5238 & 203.6770 & 51.6137 & 229 & 3.0 & 4.41 & 0.04 & 13.67 \\
UGC06541 & 173.3701 & 49.2381 & 249 & 3.0 & 4.13 & 0.23 & 13.85 \\
NGC4068 & 181.0078 & 52.5884 & 210 & 3.0 & 4.29 & 0.04 & 12.57 \\
ESO379-024 & 181.2361 & -35.7431 & 632 & 3.0 & 5.35 & 0.25 & 15.87 \\
NGC5236 & 204.2537 & -29.8656 & 513 & 3.0 & 4.79 & 0.09 & 7.50 \\
ESO381-018 & 191.1773 & -35.9666 & 624 & 3.0 & 5.32 & 0.12 & 15.06 \\
NGC4605 & 189.9975 & 61.6090 & 136 & 3.0 & 5.42 & 0.05 & 10.06 \\
PGC046885 & 201.1501 & -30.9719 & 494 & 3.0 & 4.66 & 0.13 & 15.88 \\
NGC5264 & 205.4031 & -29.9127 & 478 & 3.0 & 4.68 & 0.15 & 12.13 \\
NGC5204 & 202.4019 & 58.4193 & 201 & 3.0 & 4.71 & 0.45 & 11.30 \\
IC4316 & 205.0761 & -28.8942 & 576 & 3.0 & 4.25 & 0.08 & 14.06 \\
UGC07949 & 191.7492 & 36.4764 & 331 & 3.0 & 2.94 & 0.11 & 14.88 \\
NGC4163 & 183.0384 & 36.1694 & 171 & 3.0 & 2.92 & 0.04 & 13.29 \\
PGC039145 & 183.7421 & 36.2184 & 284 & 3.0 & 2.91 & 0.11 & 16.00 \\
PGC677373 & 204.2511 & -33.3625 & 591 & 3.0 & 4.45 & 0.31 & 17.01 \\
NGC4214 & 183.9132 & 36.3274 & 291 & 3.0 & 2.82 & 0.09 & 9.91 \\
UGC06757 & 176.7465 & 61.3351 & 82 & 3.0 & 4.51 & 0.29 & 19.38 \\
UGC08651 & 204.9742 & 40.7391 & 214 & 3.0 & 3.03 & 0.06 & 14.02 \\
NGC4190 & 183.4353 & 36.6332 & 235 & 3.0 & 2.77 & 0.08 & 13.67 \\
NGC3741 & 174.0249 & 45.2856 & 229 & 3.0 & 3.15 & 0.17 & 14.18 \\
UGC06817 & 177.7202 & 38.8807 & 251 & 3.0 & 2.59 & 0.10 & 12.59 \\
UGC08091 & 194.6678 & 14.2173 & 217 & 3.0 & 2.14 & 0.11 & 14.08 \\
\end{longtable}

\startlongtable
\begin{longtable}{lrrrrrrlr}
\caption{Dataset of 71 galaxies with TRGB-based distances from \cite{bib:Kashibadze2018} and \cite{bib:Kashibadze2020}, sorted with ascending distance from the Virgo center.\label{tab:virgo_Kashibadze}} \\
\hline
  \multicolumn{1}{c}{Name} &
  \multicolumn{1}{c}{RA} &
  \multicolumn{1}{c}{Dec} &
  \multicolumn{1}{c}{$v_\mathrm{hel}$} &
  \multicolumn{1}{c}{$\sigma_{v}$} &
  \multicolumn{1}{c}{$d_\mathrm{hel}$} &
  \multicolumn{1}{c}{$\sigma_{d}^{\dagger}$} &
  \multicolumn{1}{c}{Type} &
  \multicolumn{1}{c}{$B_T$} \\
  \multicolumn{1}{c}{} &
  \multicolumn{1}{c}{$[$deg$]$} &
  \multicolumn{1}{c}{$[$deg$]$} &
  \multicolumn{1}{c}{$[\mbox{km}\, \mbox{s}^{-1}]$} &
  \multicolumn{1}{c}{$[\mbox{km}\, \mbox{s}^{-1}]$} &
  \multicolumn{1}{c}{$[$Mpc$]$} &
  \multicolumn{1}{c}{$[$Mpc$]$} &
  \multicolumn{1}{c}{} &
  \multicolumn{1}{c}{$[$mag$]$} \\
\hline
\endfirsthead

\multicolumn{9}{c}%
{{\bfseries \tablename\ \thetable{} -- continued from previous page}} \\
\hline
  \multicolumn{1}{c}{Name} &
  \multicolumn{1}{c}{RA} &
  \multicolumn{1}{c}{Dec} &
  \multicolumn{1}{c}{$v_\mathrm{hel}$} &
  \multicolumn{1}{c}{$\sigma_{v}$} &
  \multicolumn{1}{c}{$d_\mathrm{hel}$} &
  \multicolumn{1}{c}{$\sigma_{d}^{\dagger}$} &
  \multicolumn{1}{c}{Type} &
  \multicolumn{1}{c}{$B_T$} \\
  \multicolumn{1}{c}{} &
  \multicolumn{1}{c}{$[$deg$]$} &
  \multicolumn{1}{c}{$[$deg$]$} &
  \multicolumn{1}{c}{$[\mbox{km}\, \mbox{s}^{-1}]$} &
  \multicolumn{1}{c}{$[\mbox{km}\, \mbox{s}^{-1}]$} &
  \multicolumn{1}{c}{$[$Mpc$]$} &
  \multicolumn{1}{c}{$[$Mpc$]$} &
  \multicolumn{1}{c}{} &
  \multicolumn{1}{c}{$[$mag$]$} \\
\hline
\endhead

\hline
\endfoot

\hline \hline
\multicolumn{9}{l}{$^{\dagger}$ The catalogue has no per-object distance error, $\sigma_{d} = 0.08\,d_\mathrm{hel}$ is adopted for every galaxy.}\\
\multicolumn{9}{l}{$^{\ast}$ IC3341 is a 4-$\sigma$ outlier and is excluded from the fit.}
\endlastfoot

BTS076 & 179.6833 & 27.5847 & 455 & 86.0 & 15.00 & 1.20 & I & 16.50 \\
UGC07512 = LV & 186.4217 & 2.1592 & 1506 & 2.0 & 11.86 & 0.95 & I & 15.44 \\
IC3341$^\ast$ & 186.5979 & 27.7456 & 372 & 1.0 & 11.64 & 0.93 & I & 16.69 \\
PGC043072 & 191.5637 & 10.2056 & 1142 & 1.5 & 9.64 & 0.77 & I & 15.74 \\
IC3583 & 189.1817 & 13.2589 & 1120 & 0.0 & 9.51 & 0.76 & I & 13.00 \\
GR34 & 185.5321 & 15.7992 & 1298 & 3.0 & 9.29 & 0.74 & Sm & 15.80 \\
NGC4600 & 190.0958 & 3.1178 & 852 & 34.0 & 9.29 & 0.74 & S0 & 13.27 \\
KKSG29 & 189.3088 & -10.4978 & 756 & 2.0 & 9.81 & 0.78 & I & 16.54 \\
NGC4437 = N4517 & 188.1896 & 0.1144 & 1134 & 18.0 & 8.36 & 0.67 & Sc & 11.19 \\
NGC4594 & 189.9975 & -11.6231 & 1090 & 33.0 & 9.55 & 0.76 & S0a & 9.00 \\
NGC4656 & 190.9913 & 32.1700 & 638 & 17.0 & 7.96 & 0.64 & Sm & 9.82 \\
NGC4485 & 187.6300 & 41.7008 & 480 & 17.0 & 8.91 & 0.71 & I & 12.33 \\
NGC4631 & 190.5329 & 32.5419 & 583 & 17.0 & 7.35 & 0.59 & Sd & 9.80 \\
KK149 & 187.2179 & 42.1781 & 407 & 17.0 & 8.51 & 0.68 & I & 15.00 \\
KK151 & 187.5983 & 42.9014 & 436 & 1.0 & 8.20 & 0.66 & Im & 15.84 \\
SDSSJ121811.07+465500.6 & 184.5462 & 46.9169 & 396 & 1.0 & 8.26 & 0.66 & Sm & 16.91 \\
KK144 & 186.3704 & 28.4844 & 480 & 2.0 & 5.99 & 0.48 & I & 16.07 \\
KK176 & 194.9858 & -19.4114 & 825 & 0.0 & 7.28 & 0.58 & I & 16.50 \\
NGC4258 & 184.7400 & 47.3039 & 447 & 7.0 & 7.66 & 0.61 & Sbc & 9.12 \\
AGC749241 & 190.0071 & 26.3219 & 451 & 1.0 & 5.62 & 0.45 & I & 18.30 \\
DDO120 & 185.3133 & 45.8119 & 463 & 17.0 & 7.28 & 0.58 & I & 13.55 \\
KDG101 & 184.7883 & 47.0903 & 132 & 24.0 & 7.28 & 0.58 & I & 16.11 \\
Arp211 & 189.3408 & 38.7439 & 458 & 1.0 & 6.14 & 0.49 & I & 15.70 \\
NGC4144 & 182.4938 & 46.4572 & 265 & 8.0 & 6.89 & 0.55 & Sc & 12.04 \\
DDO161 & 195.8200 & -17.4231 & 744 & 18.0 & 6.03 & 0.48 & Sm & 13.49 \\
KDG215 & 193.9187 & 19.2092 & 419 & 9.0 & 4.83 & 0.39 & I & 16.50 \\
KK177 & 195.6742 & 21.9978 & 267 & 60.0 & 4.82 & 0.39 & E & 16.62 \\
UA319 & 195.5600 & -17.2378 & 755 & 4.0 & 5.75 & 0.46 & I & 15.33 \\
DDO126 & 186.7717 & 37.1428 & 217 & 17.0 & 4.97 & 0.40 & I & 14.18 \\
NGC4395 & 186.4538 & 33.5467 & 313 & 17.0 & 4.76 & 0.38 & Sm & 10.84 \\
NGC4826 & 194.1821 & 21.6819 & 409 & 8.0 & 4.41 & 0.35 & Sa & 9.27 \\
PGC038685 & 182.4850 & 36.4350 & 327 & 0.8 & 4.85 & 0.39 & Sm & 15.45 \\
UGC07605 & 187.1612 & 35.7175 & 310 & 2.0 & 4.74 & 0.38 & I & 14.78 \\
UGCA281 & 186.5675 & 48.4939 & 279 & 0.0 & 5.70 & 0.46 & I & 15.10 \\
DDO127 & 187.1171 & 37.2347 & 275 & 17.0 & 4.72 & 0.38 & Im & 14.92 \\
LVJ1243+4127 & 190.9821 & 41.4569 & 402 & 2.0 & 4.82 & 0.39 & I & 17.20 \\
IC3687 & 190.5629 & 38.5025 & 350 & 17.0 & 4.57 & 0.37 & I & 13.75 \\
ESO443-009 & 193.7233 & -28.3408 & 645 & 1.0 & 5.97 & 0.48 & I & 17.09 \\
NGC5068 & 199.7287 & -21.0392 & 668 & 18.0 & 5.16 & 0.41 & Scd & 10.50 \\
DDO154 & 193.5225 & 27.1500 & 374 & 1.0 & 4.04 & 0.32 & I & 14.11 \\
NGC4244 & 184.3729 & 37.8078 & 247 & 17.0 & 4.31 & 0.34 & Sc & 10.44 \\
KK109 & 176.7979 & 43.6719 & 211 & 2.0 & 4.51 & 0.36 & I & 18.14 \\
UGCA292 & 189.6667 & 32.7669 & 308 & 0.1 & 3.85 & 0.31 & I & 16.05 \\
KK160 & 190.9892 & 43.6614 & 293 & 1.0 & 4.33 & 0.35 & I & 17.40 \\
NGC4449 & 187.0462 & 44.0936 & 201 & 17.0 & 4.27 & 0.34 & Sm & 10.00 \\
NGC3738 & 173.9533 & 54.5244 & 225 & 8.0 & 5.30 & 0.42 & I & 11.55 \\
UGCA365 & 204.1287 & -29.2353 & 577 & 2.0 & 5.42 & 0.43 & I & 15.48 \\
{[KKH2007]J1228+4358} & 187.1871 & 43.9717 & 226 & 16.0 & 4.07 & 0.33 & I & 15.21 \\
PGC039228 & 183.9442 & 52.3872 & 159 & 1.7 & 4.61 & 0.37 & I & 15.33 \\
ESO381-020 & 191.5017 & -33.8381 & 589 & 1.0 & 5.47 & 0.44 & I & 14.23 \\
IC4247 & 201.6850 & -30.3625 & 420 & 18.0 & 5.18 & 0.41 & I & 14.43 \\
KK195 & 200.2846 & -31.5292 & 572 & 2.0 & 5.22 & 0.42 & I & 17.17 \\
UGC06541 & 173.3704 & 49.2381 & 249 & 2.0 & 4.23 & 0.34 & I & 14.42 \\
NGC4068 & 181.0083 & 52.5892 & 210 & 2.0 & 4.39 & 0.35 & Im & 13.20 \\
NGC5236 & 204.2537 & -29.8656 & 519 & 18.0 & 4.90 & 0.39 & Sc & 8.20 \\
ESO381-018 & 191.1771 & -35.9667 & 624 & 18.0 & 5.45 & 0.44 & I & 15.77 \\
ESO444-084 & 204.3346 & -28.0447 & 587 & 3.0 & 4.61 & 0.37 & I & 15.10 \\
UGC07298 & 184.1258 & 52.2281 & 174 & 2.0 & 4.19 & 0.34 & I & 15.93 \\
KK200 & 201.1504 & -30.9719 & 494 & 2.0 & 4.76 & 0.38 & I & 15.80 \\
NGC5264 & 205.4029 & -29.9128 & 478 & 3.0 & 4.79 & 0.38 & Sm & 12.59 \\
IC4316 & 205.0762 & -28.8942 & 576 & 2.0 & 4.35 & 0.35 & I & 14.54 \\
NGC4163 & 183.0379 & 36.1686 & 162 & 5.0 & 2.99 & 0.24 & Im & 13.70 \\
HIDEEPJ1337-33 & 204.2508 & -33.3625 & 591 & 18.0 & 4.55 & 0.36 & I & 17.54 \\
NGC4214 & 183.9142 & 36.3264 & 291 & 22.0 & 2.88 & 0.23 & I & 10.21 \\
UGC06757 & 176.7467 & 61.3350 & 88 & 0.3 & 4.61 & 0.37 & Sm & 16.30 \\
NGC4190 & 183.4354 & 36.6333 & 235 & 22.0 & 2.83 & 0.23 & I & 13.85 \\
NGC3741 & 174.0250 & 45.2853 & 229 & 2.0 & 3.22 & 0.26 & I & 14.41 \\
UGC06817 & 177.7208 & 38.8803 & 251 & 1.7 & 2.65 & 0.21 & I & 13.46 \\
DDO125 & 186.9229 & 43.4958 & 206 & 2.0 & 2.61 & 0.21 & Im & 12.90 \\
GR8 & 194.6675 & 14.2175 & 217 & 2.0 & 2.19 & 0.18 & I & 14.70 \\
NGC5253 & 204.9833 & -31.6400 & 407 & 3.0 & 3.44 & 0.28 & Sm & 10.90 \\
\end{longtable}

\mbox{}

\clearpage
\end{document}